\documentclass[twocolumn, journal]{IEEEtran}

\IEEEoverridecommandlockouts
\usepackage{stfloats}
\usepackage{mathrsfs}
\usepackage{diagbox}
\usepackage{bm}
\usepackage{color}
\usepackage{graphicx}
\usepackage{amsmath}
\usepackage{amssymb}
\usepackage{algorithm}
\usepackage{algorithmic}
\usepackage{ulem}
\usepackage{lineno}
\usepackage{lettrine}
\usepackage{subfigure}
\usepackage{epstopdf}
\usepackage{stfloats}
\usepackage{color}
\usepackage{graphicx}
\usepackage{amsmath}
\usepackage{amssymb}
\usepackage{algorithm}
\usepackage{algorithmic}
\usepackage{amsmath}
\usepackage{multirow}
\usepackage{booktabs}
\usepackage{array}
\usepackage{amsthm}
\usepackage{stfloats}
\usepackage{caption}
\usepackage{subfigure}
\usepackage{bm}
\usepackage{booktabs}
\usepackage{setspace}
\usepackage{makecell}
\usepackage{cases}

\newcommand{\be}{\begin{equation}}
\newcommand{\ee}{\end{equation}}
\newcommand{\bea}{\begin{eqnarray}}
\newcommand{\eea}{\end{eqnarray}}
\newcommand{\ba}{\begin{array}}
\newcommand{\ea}{\end{array}}
\newcommand{\nid}{\noindent}

\allowdisplaybreaks[4]

\title{Sparsity Exploitation via Joint Receive Processing and Transmit Beamforming Design for MIMO-OFDM ISAC Systems
	\thanks{Z. Xiao, M. Li, and W. Wang are with the School of Information and Communication
		Engineering, Dalian University of Technology, Dalian 116024, China (e-mail:
		xiaozichao@mail.dlut.edu.cn; mli@dlut.edu.cn; wangwei2023@dlut.edu.cn).}
    \thanks{R. Liu is with the Center for Pervasive Communications and Computing, University of California, Irvine, CA 92697, USA (e-mail: rangl2@uci.edu).}
	\thanks{Q. Liu is with the School of Computer Science and Technology, Dalian University
		of Technology, Dalian 116024, China (e-mail: qianliu@dlut.edu.cn).}}

\author{Zichao Xiao,
		Rang Liu,~\IEEEmembership{Graduate Student Member,~IEEE,}
        Ming Li,~\IEEEmembership{Senior Member,~IEEE,}\\
        Wei Wang,~\IEEEmembership{Member,~IEEE}, 
		and Qian Liu,~\IEEEmembership{Member,~IEEE}
}

\begin{document}
\maketitle
\thispagestyle{empty}
\begin{abstract}
Integrated sensing and communication (ISAC) is widely recognized as a pivotal enabling technique for the advancement of future wireless networks. This paper aims to efficiently exploit the inherent sparsity of echo signals for the multi-input-multi-output (MIMO) orthogonal frequency division multiplexing (OFDM) based ISAC system. A novel joint receive echo processing and transmit beamforming design is presented to achieve this goal. Specifically, we first propose a compressive sensing (CS)-assisted estimation approach to facilitate ISAC receive echo processing, which can not only enable accurate recovery of target information, but also allow substantial reduction in the number of sensing subcarriers to be sampled and processed. Then, based on the proposed CS-assisted processing method, the associated transmit beamforming design is formulated with the objective of maximizing the sum-rate of multiuser communications while satisfying the transmit power budget and ensuring the received signal-to-noise ratio (SNR) for the designated sensing subcarriers. In order to address the formulated non-convex problem involving high-dimensional variables, an effective iterative algorithm employing majorization minimization (MM), fractional programming (FP), and the nonlinear equality alternative direction method of multipliers (neADMM) with closed-form solutions has been developed. Finally, extensive numerical simulations are conducted to verify the effectiveness of the proposed algorithm and the superior performance of the introduced sparsity exploitation strategy.
\end{abstract}

\begin{IEEEkeywords}
Integrated sensing and communications (ISAC), sparsity, estimation, beamforming, compressive sensing (CS).
\end{IEEEkeywords}

\section{Introduction}

Integrated sensing and communication (ISAC), which focuses on performing both sensing and communication functionalities in one system via sharing the same transmitted waveform and a majority of hardware, enjoys high spectrum/hardware/power efficiency and has various promising application prospects, such as intelligent transportation, manufacturing, healthcare, etc. ISAC is expected to enable the next-generation mobile networks to be ubiquitous sensing ability while providing simultaneous communication service.
Owing to these attractive superiorities, ISAC is receiving increasing research interest and has already been acknowledged as a critical enabling technology for sixth generation (6G) wireless networks \cite{ITU future prospective}-\cite{F Liu overview}. 

Many types of ISAC systems have been proposed, which can be generally divided into three categories: Radar system based, communication system based, and dedicated designed dual-functional system. 
Specifically, the radar system based ISAC usually incorporates communication modulation and demodulation methods within a standard radar waveform, e.g., linear frequency modulated signals \cite{Chirp based 1}, frequency-hopping (FH) signals \cite{FH radar based 2}, frequency agile radar \cite{FAR radar based}, etc. The communication system based ISAC aims to realize sensing function using a standard communication waveform, especially the OFDM signal which has been widely employed in mobile and WiFi networks \cite{C Sturm}, \cite{80211 1}.
As for the third type, there is no clear boundary between it and the previous two categories, but it undoubtedly offers more design flexibility. For instance, the ISAC systems employ the newly emerged orthogonal time frequency space (OTFS) waveform \cite{OTFS ISAC} or reconfigurable intelligent surface (RIS) \cite{RIS ISAC}.
The widespread implementation of mobile and WiFi communication systems has led to the recognition of OFDM ISAC as a highly promising technology for enabling future wireless networks with pervasive sensing capabilities.
Therefore, this paper focuses on the investigation of the design and optimization of the ISAC system with a particular emphasis on the well-studied multi-input multi-output (MIMO) OFDM waveform and transceiver architecture.

Two prominent issues in the design of MIMO-OFDM ISAC systems are the design of transmit beamforming and the processing of received echo signals. The design of ISAC transmit beamforming involves optimizing the precoding matrix to provide a dual-functional beampattern that exhibits desirable characteristics in terms of both sensing and communication design metrics, e.g., radar beampattern similarity and spectral efficiency of communication \cite{beampattern error and rate}, radar beampattern similarity and communication multi-user interference (MUI) \cite{beampattern error and MUI}, radar signal-to-interference-plus-noise ratio (SINR) and communication SINR \cite{SINR and SINR 1}, \cite{SINR and SINR 2}, radar SINR and communication quality-of-service (QoS) \cite{SINR and QoS}. 
Meanwhile, the research of received echo signal processing focuses on target parameter estimation, e.g., angle, range, and Doppler, using MIMO-OFDM communication waveforms. 
The authors in \cite{CS-ISAC Rahman1} and \cite{CS-ISAC Rahman2} proposed two block compressive sensing (CS) based algorithms, including high-complexity general direct sensing for data signal part and low-complexity indirect sensing methods applicable for orthogonal training/pilot part. 
Afterward, the authors in \cite{MIMO OFDM ISAC successive estimation 1} and \cite{MIMO OFDM ISAC successive estimation 2} presented a more efficient successive estimation method.
This approach begins with an estimation of the angle, which is subsequently employed for extracting the information of range and Doppler.   
More recently, the authors in \cite{our estimation work} introduced a novel joint estimation method that achieves better sensing performance in MIMO-OFDM ISAC systems. This strategy involves jointly estimating the angle-range-Doppler by effectively exploiting all the received echo signals within a coherent processing interval (CPI).

However, despite extensive research efforts on MIMO-OFDM ISAC designs, the exploitation of sparsity has not received sufficient attention. The echo signals usually exhibit intrinsic sparsity in most sensing scenarios, since they are characterized by limited occupied angle-distance-velocity cells that include items of interest.  
According to the well-known CS technique, a sparse signal can be accurately recovered from significantly fewer samples than what is mandated by the Nyquist-Shannon sampling theorem \cite{CS theory 1}. 
This results in a reduction of system overhead or an enhancement of system performance. 
In the field of radar, researchers have utilized sparsity to reduce the necessary number of antennas or radar pulses \cite{CS aided sensing (reduce cost)1}, \cite{CS aided sensing (reduce cost)3}, achieve better sensing resolution \cite{CS aided sensing (enhance resolution)1}, and improve the target detection performance \cite{CS aided sensing (enhance detection)}, etc.
In the field of communication, the sparsity inherent in the millimeter wave (mmWave) or terahertz (THz) channel has also been utilized to decrease the pilot overhead and enhance the accuracy of channel estimation \cite{CS aided communications1}, \cite{CS aided communications2}.

Unlike the aforementioned two aspects, sparsity exploitation in ISAC systems is extremely challenging due to the more sophisticated signal form. Only a limited number of studies have been conducted in this area.
In the works \cite{CS-ISAC Rahman1}, \cite{CS-ISAC Rahman2}, the authors utilized the multi-measurement vector CS algorithm as an effective estimation tool for processing complicated MIMO-OFDM echo signals, but did not intentionally exploit the sparsity in an effort to further improve ISAC system efficiency.
To ensure the construction of a virtual array for achieving high angular resolution while allocating as many subcarriers as possible for communications, the authors in \cite{MIMO OFDM ISAC successive estimation 2} proposed a novel shared and private subcarrier strategy for MIMO-OFDM ISAC systems, in which the CS technique is employed to solve the relevant angle estimation.
In addition to the OFDM waveform, the authors in \cite{CS aided ISAC} provided a CS perspective for mmWave massive MIMO ISAC, in which the CS technique is employed for target sensing with energy-efficient widely spaced array and channel estimation for hybrid digital-analog transceiver.
Although the effectiveness in minimizing the pilot overhead of this approach has been demonstrated, its direct applicability to MIMO-OFDM ISAC systems may not be easy, and the data payload sparsity has not been considered either.
Therefore, it is still necessary to further study the sparsity exploitation from both the transmitter and receiver sides of the MIMO-OFDM ISAC system.

Motivated by these findings, in this paper we aim to investigate the sparsity exploitation via joint receive signal processing and transmit beamforming design to further improve the efficiency of MIMO-OFDM ISAC systems.
The main contributions can be summarized as follows:

\begin{itemize}
\item Firstly, we propose a novel CS-based estimation method for MIMO-OFDM ISAC systems. This method provides a feasible sparsity exploitation strategy to reduce sensing expenses by effectively leveraging the characteristics of abundant OFDM subcarriers.
In contrast to prior research \cite{our estimation work}, this novel approach enables the utilization of echo signals of a reduced number of subcarriers for accomplishing the same estimation performance. Consequently, this advancement facilitates the resource allocation of MIMO-OFDM ISAC systems in the following transmit beamforming design.

\item Then, based on the proposed estimation approach, we turn to investigate the associated transmit beamforming design to further enhance the sparsity exploitation.
Particularly, we utilize a novel sensing metric that quantifies the signal-to-noise ratio (SNR) of the echo signals of the subcarriers used for the estimation procedure.
Afterward, a beamforming design problem, which aims at maximizing the communication sum-rate while guaranteeing the sensing requirement and total transmit power budget, is formulated.
Unlike the beamforming designs for conventional MIMO-OFDM communication/radar-only systems, the considered ISAC design encounters highly coupled constraints for the precoders of all the subcarriers, making this high-dimensional non-convex problem challenging.

\item Toward the formulated high-dimensional non-convex problem, we first employ majorization-minimization (MM) and fractional-programming (FP) methods to convert it into a series of convex problems.
Then, the novel nonlinear equality constrained alternative direction method of multipliers (neADMM) approach is exploited to convert the problem into multiple low-dimensional sub-problems with tractable forms.
An efficient optimal closed-form solution is derived in detail for each sub-problem.

\item Extensive simulation results are provided to verify the effectiveness of the proposed algorithm and the advantages of the proposed sparsity exploitation strategy.
The system efficiency is dramatically enhanced by employing the joint CS-based estimation and the associated transmit beamforming design.
\end{itemize}

\textit{Notation}:
Lower-case, boldface lower-case, and upper-case letters indicate scalars, column vectors, and matrices, respectively.
$(\cdot)^T$, $(\cdot)^*$, and $(\cdot)^H$  denote the transpose, conjugate, and transpose-conjugate operations, respectively.
$\mathbb{E} \{ \cdot \}$ represents statistical expectation.
$| a |$, $\| \mathbf{a} \|$, and $\| \mathbf{A} \|$ are the magnitude of a scalar $a$, 2-norm of a vector $\mathbf{a}$, and the Frobenius norm of a matrix $\mathbf{A}$, respectively.
$\mathbb{C}$ and $\mathbb{R}$ denote the set of complex and real numbers, respectively.
$\mathrm{Rank}(\mathbf{A})$ and $\text{Tr}(\mathbf{A})$ returns the rank and trace of matrix $\mathbf{A}$, respectively.
$\Re\{ \cdot \}$ and $\Im\{ \cdot \}$ extract the real and image parts of a complex number, respectively.
$\sqrt{a}$ returns the square root of scalar $a$.
$\mathbf{I}_{N}$ indicates an $N \times N$ identity matrix.
$\otimes$ denotes the Kronecker product operation.
$\mathbf{A}(i,j)$ denotes the element of the $i$-th row and the $j$-th column of matrix $\mathbf{A}$.
$\mathrm{Diag}\{\mathbf{a}\}$ denotes a square diagonal matrix with vector $\mathbf{a}$ as the main diagonal.
$\mathrm{vec}(\mathbf{A})$ vectorizes the matrix $\mathbf{A}$.
$\mathbf{A}^\dag$ returns the Moore-Penrose pseudoinverse of matrix $\mathbf{A}$.
$\mathbf{A}\setminus \mathbf{0}^T$ denotes removing all the row vectors whose elements are only 0 of matrix $\mathbf{A}$.
\section{System Model}
We consider a mono-static MIMO-OFDM ISAC system, where a dual-functional base station (BS) equipped with two separate uniform linear arrays (ULAs) of $N_\mathrm{t}$ transmit antennas and $N_\mathrm{r}$ receive antennas performs active sensing and downlink communications.
Specifically, the BS simultaneously serves $K$ downlink single-antenna communication users and estimates the angle-distance-velocity information of targets located in an interested region with given directions and coverage ranges.
We confine ourselves to a scenario in which the number of users is limited by $N_\mathrm{t}$, i.e., $K \leq N_\mathrm{t}$, which is widely accepted to allow the BS to be more capable of handling harmful MUI.

\subsection{Transmitted Signal}

In the considered MIMO-OFDM ISAC system, the carrier frequency is set as $f_\mathrm{c}$.
There are a total of $N_\mathrm{s}$-subcarriers with frequency spacing $\Delta f$, and the OFDM symbol duration is set as $T_\mathrm{d}\triangleq{1}/{\Delta f}$.
Additionally, the inserted cyclic prefix (CP) for avoiding inter symbol interference (ISI) at each OFDM symbol is of $T_\mathrm{cp}$ long, hence the total symbol duration is $T\triangleq T_\mathrm{d}+T_\mathrm{cp}$.
In each subcarrier, the communication modulated symbol vector of $K$ users is denoted as $\mathbf{s}_i[l]\in \mathbb{C}^{ K \times 1}$, where $i=0,\ldots, N_\mathrm{s}-1$ denotes the subcarrier index, $l=0,\ldots,L-1$ is the symbol slot, and $L$ is the signal frame and sensing CPI length.
Each element of $\mathbf{s}_i[l]$ is independently selected from the quadrature amplitude modulation (QAM) constellation of unit power and we have $\mathbb{E}\{\mathbf{s}_i[l]\mathbf{s}_i[l]^H\}=\mathbf{I}_K$.

In order to provide a favorable beampattern for sensing-communication, dual-information symbols are firstly precoded by beamforming matrices $\mathbf{W}_i \triangleq [\mathbf{w}_{i,1},\ldots,\mathbf{w}_{i,K}] \in \mathbb{C}^{ N_\mathrm{t} \times K}$, $\forall i$, which will be elaborately designed in Sec. \ref{sec: beamforming}.
Correspondingly, the $l$-th transmitted OFDM baseband symbols on the $i$-th subcarrier can be represented as
\begin{equation} \label{precoding process}
\mathbf{x}_i[l] = \mathbf{W}_i \mathbf{s}_i[l].
\end{equation}


\subsection{Communication Model}
At the communication receiver, after a series of processing operations \cite{our estimation work}, the frequency-domain signal of the $k$-th user on the $i$-th subcarrier can be modeled as
\begin{equation}\label{communication receive signal}
y_{i,k}[l] = \mathbf{h}_{i,k}^H \mathbf{W}_i \mathbf{s}_i[l]+z_{i,k}[l] , ~~\forall i,~ k,
\end{equation}
where $\mathbf{h}_{i,k}\in \mathbb{C}^{N_\mathrm{t}\times 1}$ denotes the frequency domain channel and $z_{i,k} \in \mathcal{CN}(0,\sigma_\mathrm{c}^2)$ denotes the independent and identically distributed (i.i.d.) additive white Gaussian noise (AWGN).
Consequently, the sum-rate of all the $K$ users on all the subcarriers can be calculated as
\begin{equation} \label{sum rate}
\begin{aligned}
R_\mathrm{c} = \sum_{i=0}^{N_\mathrm{s}-1} \sum_{k=1}^K \log\left(1+  \frac{|\mathbf{h}_{i,k}^H\mathbf{w}_{i,k}|^2}{\sum_{j\neq k} |\mathbf{h}_{i,k}^H\mathbf{w}_{i,j}|^2 + \sigma_\mathrm{c}^2 }  \right).
\end{aligned}
\end{equation}

To focus on the beamforming design of the considered ISAC system, the following two assumptions are considered: 
(A1) The exact and instantaneous channel state information (CSI), i.e., $\mathbf{h}_{i,k}$, $\forall i,k$, is available at the BS.
(A2) The channel matrix $\mathbf{H}_i\triangleq [\mathbf{h}_{i,1},\ldots,\mathbf{h}_{i,K}]$ has full collum rank, i.e., $\mathrm{Rank}(\mathbf{H}_i) =K$, $\forall i$.
As for (A1), many existing works have provided efficient channel estimation methods for the considered communication systems \cite{channel estimation 2}.
As for (A2), it is typically satisfied in the considered partially loaded system \cite{book: Emil}.

\subsection{Sensing Model}

We assume a total of $Q$ point targets at the angle $\theta_q$ within distance $d_q$ and velocity $v_q$ relative to the BS, $q=1,\ldots,Q$.
According to the derivations in our previous work \cite{our estimation work}, the frequency-domain received echo on the $m$-th receive antenna related to the $i$-th subcarrier is presented as
\begin{align}
    y(m,i,l)
    \triangleq
    \sum_{q=1}^Q \beta_q &\sqrt{\mathrm{PL}(2d_q)}  \mathbf{a}\left(\omega_\mathrm{t}(\theta_q)\right)^H\mathbf{x}_i[l]
    \label{established_echo_model}\\
    &e^{\jmath m\omega_\mathrm{r}(\theta_q) }
    e^{\jmath i\omega_\mathrm{d}(d_q)}
    e^{\jmath l\omega_\mathrm{v}(v_q) }
       + z(m,i,l), \nonumber
\end{align}
where $m=0,\ldots,N_\mathrm{r}-1$, $\beta_q$ denotes the reflection coefficient of the $q$-th target with $ \mathbb{E}\{|\beta_q|^2\} =\sigma_\beta^2$, $\mathrm{PL}(d)\triangleq c_\mathrm{ref}(\frac{d}{d_\mathrm{ref}})^{-\alpha}$ denotes the distance-dependent path loss, where $c_\mathrm{ref}$ is the loss for the reference distance $d_\mathrm{ref}$, $d$ is the path distance, and $\alpha$ is the path-loss exponent.
$\mathbf{a}(\omega)\triangleq [e^{\jmath0},e^{\jmath1\omega},\ldots,e^{\jmath(N_\mathrm{t}-1)\omega}]^T$ represents the transmit steering vector of frequency $\omega$,
$z(m,i,l) \sim \mathcal{CN}(0,\sigma_\mathrm{s}^2)$ is the AWGN.
$\omega_\mathrm{t}(\theta_q)$, $\omega_\mathrm{r}(\theta_q)$, $\omega_\mathrm{d}(d_q)$, and $\omega_\mathrm{v}(v_q)$ denote the angle-dependent transmit, angle-dependent receive, distance-dependent, and velocity-dependent digital frequency, respectively, which are defined as follows
\begin{subequations}
\begin{align}
\omega_\mathrm{t}(\theta_q)&\triangleq 2\pi  d_\mathrm{t} \sin\theta_q f_\mathrm{c}/c,  \label{transmit spatial frequency} \\
\omega_\mathrm{r}(\theta_q)&\triangleq -2\pi  d_\mathrm{r}\sin\theta_q f_\mathrm{c}/c,\label{receive spatial frequency}\\
\omega_\mathrm{d}(d_q)&\triangleq -2\pi \Delta f \frac{2d_q}{c},\label{distance dependent frequency}\\
\omega_\mathrm{v}(v_q)&\triangleq 4\pi T v_q f_\mathrm{c}/c, \label{velocity dependent frequency}
\end{align}
\end{subequations}
where $c$ is the speed of light, $d_\mathrm{t}$ and $d_\mathrm{r}$ are the transmit and receive antenna spacings, respectively.

\section{Compressive Sensing Assisted \\Target Parameter Estimation Approach} \label{sec: CS-estimation}

In this section, we are firstly committed to mathematically analyzing the sparsity in the discrete Fourier transform (DFT) based estimation method proposed in previous work \cite{our estimation work}.
Subsequently, a novel estimation approach is introduced, leveraging CS techniques, that achieves comparable performance to the conventional DFT-based algorithm while utilizing a reduced number of OFDM subcarriers.  
The implementation of this novel estimation approach associated with the beamforming designs presented in the next section can enable the ISAC system to allocate more resources for communication functions and achieve superior communication performance.  

\subsection{Sparsity Identification}
In our previous work \cite{our estimation work}, we propose a DFT-based joint angle-range-velocity estimation method for MIMO-OFDM ISAC systems.
In order to better utilize the CS technique to enhance the estimation performance by exploiting sparsity, we focus on analyzing the sparsity contained in the estimation process in this subsection.
The estimation procedures in \cite{our estimation work} can be briefly summarized as follows.
Firstly, $N_\mathrm{a}$-point ($N_\mathrm{a}\geq N_\mathrm{r}$) DFT is applied on all the received echoes (\ref{established_echo_model}) along the spatial dimension, i.e.,
\be
\label{spatial DFT}
Y_{i,l}(n_{\mathrm{a}})
	\triangleq \frac{1}{N_\mathrm{r}}\sum_{m=0}^{N_\mathrm{r}-1}y(m,i,l) e^{-\jmath m \frac{2\pi n_\mathrm{a}}{N_\mathrm{a}}},
\ee
where $n_{\mathrm{a}}=-\lfloor N_\mathrm{a}/2\rfloor,\ldots,\lfloor N_\mathrm{a}/2\rfloor-1$.
Then, element-wise division for removing the signal-dependent coefficients is operated as
\be \label{MS remove}
y_{i,l}(n_{\mathrm{a}})\triangleq Y_{i,l}(n_{\mathrm{a}})/(\alpha_{n_{\mathrm{a}}}\mathbf{a}(-\frac{d_\mathrm{t}n_{\mathrm{a}}2\pi}{d_\mathrm{r}N_\mathrm{a}})^H\!\mathbf{x}_i[l])
, ~\forall i,~l,~n_{\mathrm{a}},
\ee
with a scaling factor
\be
\alpha_{n_{\mathrm{a}}}
=
\sqrt{
\frac
{\sum_{i,l}
\left|\frac{Y_{i,l}(n_\mathrm{a})}
{\mathbf{a}(-\frac{d_\mathrm{t}n_{\mathrm{a}}2\pi}{d_\mathrm{r}N_\mathrm{a}})^H\mathbf{x}_i[l]
}\right|^2 }
{\sum_{i,l}|Y_{i,l}(n_\mathrm{a})|^2}
}.
\ee
After that, $N_\mathrm{v}$-point ($N_\mathrm{v}\geq L$) DFT is applied on the obtained data (\ref{MS remove}) along the Doppler dimension as
\begin{align}  \label{slow time DFT}
Y_{i}(n_{\mathrm{a}},n_{\mathrm{v}})
	\triangleq \frac{1}{L}\sum_{l=0}^{L-1}y_{i,l}(n_{\mathrm{a}}) e^{-\jmath l \frac{2\pi n_\mathrm{v}}{N_\mathrm{v}}},
\end{align}
where $n_{\mathrm{v}}=-\lfloor N_\mathrm{v}/2\rfloor,\ldots,\lfloor N_\mathrm{v}/2\rfloor-1$.
Then, $N_\mathrm{d}$-point ($N_\mathrm{d}\geq N_\mathrm{s}$) DFT is applied along the delay dimension as
\begin{align} \label{fast time DFT}
Y(n_{\mathrm{a}},n_{\mathrm{d}},n_{\mathrm{v}})
	\triangleq \frac{1}{N_\mathrm{s}}\sum_{i=0}^{N_\mathrm{s}-1} Y_{i}(n_{\mathrm{a}},n_{\mathrm{v}}) e^{-\jmath i \frac{2\pi n_\mathrm{d}}{N_\mathrm{d}}},
\end{align}
where $n_{\mathrm{d}}=- N_\mathrm{d}+1,\ldots,0$, $\forall n_{\mathrm{a}},n_{\mathrm{v}}$.
Operation (\ref{fast time DFT}) can be equivalently written in a compact form as
\be
\begin{aligned} \label{sparse signal}
\mathbf{y}(n_{\mathrm{a}},n_{\mathrm{v}})&\triangleq
[Y(n_{\mathrm{a}},-N_\mathrm{d}+1,n_{\mathrm{v}}),\ldots,Y(n_{\mathrm{a}},0,n_{\mathrm{v}}) ]^T\\
&=\mathbf{F}_{N_\mathrm{d}}\widehat{\mathbf{y}}(n_{\mathrm{a}},n_{\mathrm{v}}),
\end{aligned}
\ee
where $\mathbf{F}_{N_\mathrm{d}} \in \mathbb{C}^{N_\mathrm{d} \times N_\mathrm{d}}$ is the DFT matrix whose $(m,n)$-th element is defined as $\mathbf{F}_{N_\mathrm{d}}(m,n) =
    e^{-\jmath (n-1) \frac{ (m-N_\mathrm{d}) 2\pi}{N_\mathrm{d}}  } $, $
\widehat{\mathbf{y}}(n_{\mathrm{a}},n_{\mathrm{v}})\triangleq [Y_{0}(n_{\mathrm{a}},n_{\mathrm{v}}),\ldots,Y_{N_\mathrm{s}-1}(n_{\mathrm{a}},n_{\mathrm{v}}),0,\ldots,0]_{1\times N_\mathrm{d}}^T
$.
Finally, the estimated parameter information can be acquired by identifying all the peaks that appeared in the processed data (\ref{sparse signal}), as elaborated in \cite{our estimation work}.

According to the findings and simulations presented in our previous study \cite{our estimation work}, there exists at most $Q$ peaks observing from the final vector (\ref{sparse signal}) subjected to AWGN.
Therefore, when the number of peaks is much smaller than its length, i.e., $Q \ll N_\mathrm{d}$, vector (\ref{sparse signal}) is \textit{sparse} (rigorously speaking, it is \textit{compressible signal} whose entries decay rapidly when sorted in order
of decreasing magnitude) \cite{CS optimization}.
The condition of sparsity is consistently met as a result of the small number of interested targets and the sufficiently large number of OFDM carriers.

\subsection{Compressed Sensing Assisted Estimation} \label{CS aided estimation method}
According to the previous discussion, sparse vector (\ref{sparse signal}) is obtained by DFT analysis with $N_\mathrm{s}$-effective-point sampled signal $\widehat{\mathbf{y}}(n_{\mathrm{a}},n_{\mathrm{v}})$.
Supported by CS theory \cite{CS theory 1}, the sparsity of a signal can be exploited to recover it from far fewer samples than required by the Nyquist-Shannon sampling theorem.
Therefore, in this subsection, we leverage the CS technique to replace the DFT operation by utilizing fewer sampled points, thereby reducing the number of subcarriers required for sensing purposes.

Specifically, only $N_\mathrm{sel}$ effective samples are selected from $\widehat{\mathbf{y}}(n_{\mathrm{a}},n_{\mathrm{v}})$ for obtaining the sparse signal $\mathbf{y}(n_{\mathrm{a}},n_{\mathrm{v}})$,  $N_\mathrm{sel}\leq N_\mathrm{s}$. And the previous recovery equation (\ref{sparse signal}) is reformulated as
\be
\begin{aligned} \label{sparse recovery}
\mathbf{\Phi}(N_\mathrm{sel})\mathbf{F}_{N_\mathrm{d}}^{-1}\mathbf{y}(n_{\mathrm{a}},n_{\mathrm{v}})
=\mathbf{\Phi}(N_\mathrm{sel})\widehat{\mathbf{y}}(n_{\mathrm{a}},n_{\mathrm{v}}),
\end{aligned}
\ee
where the preset selection matrix is defined as $\mathbf{\Phi}(N_\mathrm{sel})\triangleq\mathrm{Diag}\{ [\phi_0,\ldots,\phi_{N_\mathrm{s}-1} ,1,\ldots,1 ]_{1\times N_\mathrm{d}}^T \}\backslash\mathbf{0}^T$ with $\phi_i\in\{0,1\}$ and $\sum_i \phi_i=N_\mathrm{sel}$.
With the introduction of selection matrix $\mathbf{\Phi}(N_\mathrm{sel})$ for reducing the required sampling point, the recovery equation (\ref{sparse recovery}) becomes underdetermined, which has no solution or infinite solutions via direct equation solving.
Thanks to the sparsity of $\mathbf{y}(n_{\mathrm{a}},n_{\mathrm{v}})$ and the CS theory, a satisfactory recovery can be realized by various algorithms \cite{CS optimization}, e.g., orthogonal matching pursuit (OMP), convex optimization based method, etc.
Here, we take the widely adopted $\ell_1$-norm method as an example
\begin{subequations}\label{CS problem}
\begin{align}
&\min \limits_{\mathbf{y}(n_{\mathrm{a}},n_{\mathrm{v}})} ~~ \|\mathbf{y}(n_{\mathrm{a}},n_{\mathrm{v}})\|_1 \\
&\text{s.t.}~~ \mathbf{\Phi}(N_\mathrm{sel})\mathbf{F}_{N_\mathrm{d}}^{-1}\mathbf{y}(n_{\mathrm{a}},n_{\mathrm{v}})
=\mathbf{\Phi}(N_\mathrm{sel})\widehat{\mathbf{y}}(n_{\mathrm{a}},n_{\mathrm{v}}).
\end{align}
\end{subequations}
Sparse signal recovery problem (\ref{CS problem}) is convex and can be directly solved by the existing convex toolbox, e.g., CVX.

Our proposed CS-aided estimation method is established via replacing the DFT operation (\ref{sparse signal}) with optimization (\ref{CS problem}). 
It should be noted that there are rigorous prerequisites that indicate what quantity of selected samples $N_\mathrm{sel}$ and what form of matrix $\mathbf{\Phi}(N_\mathrm{sel})$ is more suitable for recovering sparse signal without distortion \cite{CS RIP 1}, \cite{CS RIP 2}.
In this paper we focus on applying CS for the sparsity exploitation of MIMO-OFDM ISAC systems and the selection matrix $\mathbf{\Phi}(N_\mathrm{sel})$ is randomly constructed with a given $N_\mathrm{sel}$.

This CS-aided estimation method requires the processing of echo signals that have been sampled from a significantly reduced number of OFDM subcarriers.
Therefore, this approach exhibits the potential to enable ISAC systems to achieve a more adaptable allocation of OFDM resources, thereby effectively balancing the performance demands of sensing and communication. 
In the next section, we present a transmit beamforming design algorithm associated with this CS-aided estimation method, that focuses solely on ensuring the sensing receiving quality of the selected subcarriers as well as enhancing communication performance.

\section{Associated Transmit Beamforming Design} \label{sec: beamforming}

In this section, a beamforming design problem, which is based on the proposed CS-aided estimation method and aims at maximizing the communication sum-rate while ensuring the target estimation performance, is formulated. Toward the formulated high-dimensional non-convex problem, an efficient MM-FP-neADMM based iterative algorithm with closed-form solutions is then proposed.

\subsection{Problem Formulation}
In the proposed CS-aided estimation method, all the received echo signals of selected subcarriers $(\phi_i=1, \forall i)$ in one CPI are utilized.
Therefore, according to the echo model (\ref{established_echo_model}), the SNR of the received echo signal from a target with angle $\theta$ and range $d$ can be calculated as\vspace{-2ex}

\begin{small}
\begin{align}
\mathrm{SNR}_{\mathrm{r}}(\theta,d,&\mathbf{\Phi}(N_\mathrm{sel}))
\!\triangleq\!  \frac{\sigma_\beta^2\mathrm{PL}(2d)\mathbb{E}\{\sum_{m,i,l}\phi_i|  \mathbf{a}\left(\omega_\mathrm{t}(\theta)\right)^H\mathbf{x}_i[l]|^2\}}{\mathbb{E}\{\sum_{m,i,l}\phi_i|z(m,i,l)|^2\}} \nonumber\\
=&\frac{\sigma_\beta^2 \mathrm{PL}(2d)\sum_i\phi_i\mathbf{a}\left(\omega_\mathrm{t}(\theta)\right)^H \mathbf{W}_i \mathbf{W}_i^H\mathbf{a}\left(\omega_\mathrm{t}(\theta)\right)}
    {N_\mathrm{sel}\sigma_\mathrm{s}^2}. \label{received SNR}
\end{align}
\end{small} \vspace{-2ex}

Since the estimation performance is positively related to the received echo SNR \cite{our estimation work}, in this paper we aim to design the beamforming matrix $\mathbf{W}_i$, $\forall i$ to maximize the communication sum-rate while assuring the sensing SNR requirement and the transmit power budget.
The beamforming design problem is thus formulated as
\begin{subequations} \label{beamforming problem}
\begin{align}
\max\limits_{\mathbf{W}_i,\forall i}  &~~R_\mathrm{c} \label{objective function}\\
\text{s.t.} ~ &~~\mathrm{SNR}_{\mathrm{r}}(\theta_g,d_0,\mathbf{\Phi}(N_\mathrm{sel})) \geq  \Gamma_0,~~\forall g, \label{SNR constraint}\\
& ~\sum_{i} \|\mathbf{W}_i\|^2 \leq P_0, \label{power constraint}
\end{align}
\end{subequations}
where $\theta_g$, $g=1,\ldots,G$, is the preset sensing directions of interest, $d_0$ denotes the maximum sensing range of interest, $\Gamma_0$ is the sensing received SNR threshold, and $P_0$ is the transmit power budget.
Regarding the property that the directional beam emitted by antenna arrays will always has a main-lobe of a certain width, with a proper setting of $\theta_g$, the sensing performance of a continuous angle range can be assured.
In this paper, $\theta_g$ is generated as $\theta_g=\theta_\mathrm{a}+(g-1)\frac{\theta_\mathrm{b}-\theta_\mathrm{a}}{G-1}$ to ensure the estimation performance of interested region within angular direction $\theta_\mathrm{a}\sim \theta_\mathrm{b}$.

It can be seen that the optimization problem (\ref{beamforming problem}) is a non-convex problem due to the multiple logarithmic functions of fractional terms in the objective function (\ref{objective function}) and the non-convex quadratic constraint (\ref{SNR constraint}) with the $N_\mathrm{s}\times N_\mathrm{t}\times K$-dimensional variable, which greatly hinders finding a straightforward solution.
In order to solve these difficulties, in the next subsection, we first utilize the MM, FP, and neADMM methods to convert the original problem into a series of low-dimensional sub-problems, and then develop efficient closed-form solutions to iteratively solve them.

\subsection{MM-FP-neADMM Algorithm Design}
In order to tackle the high-dimensional non-convex problem (\ref{beamforming problem}), MM and FP methods are first respectively applied to transform the non-convex quadratic constraint (\ref{SNR constraint}) and complicated objective function (\ref{objective function}) into a series of convex forms.
Then, the neADMM framework with introducing auxiliary variables is utilized to transform these high-dimensional convex problems into multiple low-dimensional sub-problems.
Finally, elaborated closed-form solutions are provided for solving these sub-problems.
The details of the algorithm development are described as follows.

\subsubsection{MM Transformation of Constraint (\ref{SNR constraint})}
Firstly, the non-convex quadratic constraints (\ref{SNR constraint}) can be iteratively approximated with a series of tractable lower-bounded surrogate constraints according to MM method \cite{MM 2}.
The procedure for deriving the surrogate constraint is described below.

Using some basic linear algebra laws and first-order Taylor expansion, the left term of constraint (\ref{SNR constraint}) at point ${\mathbf{w}_{i,k}^{(t)}}$ can be lower-bounded as\vspace{-2ex}

\begin{small}
\begin{subequations}\label{MM transform}
\begin{align}
\mathbf{a}&\left(\omega_\mathrm{t}(\theta_g)\right)^H \mathbf{W}_i \mathbf{W}_i^H\mathbf{a}\left(\omega_\mathrm{t}(\theta_g)\right)\\
&\overset{(a)}{=}\mathrm{Tr}\{\mathbf{W}_i^H\mathbf{a}(\omega_\mathrm{t}(\theta_g))\mathbf{a}(\omega_\mathrm{t}(\theta_g))^H \mathbf{W}_i\}\\
&\overset{(b)}{=}\sum_{k=1}^K \mathbf{w}_{i,k}^H \mathbf{a}(\omega_\mathrm{t}(\theta_g))\mathbf{a}(\omega_\mathrm{t}(\theta_g))^H \mathbf{w}_{i,k}\\
&\overset{(c)}{\geq} \sum_{k=1}^K  \bigg(
{\mathbf{w}_{i,k}^{(t)}}^H \mathbf{a}(\omega_\mathrm{t}(\theta_g))\mathbf{a}(\omega_\mathrm{t}(\theta_g))^H {\mathbf{w}_{i,k}^{(t)}}   \\
&~~~~~~~~~~~~~+ (\mathbf{w}_{i,k}-{\mathbf{w}_{i,k}^{(t)}})^H \mathbf{a}(\omega_\mathrm{t}(\theta_g))\mathbf{a}(\omega_\mathrm{t}(\theta_g))^H{\mathbf{w}_{i,k}^{(t)}}  \nonumber \\
&~~~~~~~~~~~~~+ {\mathbf{w}_{i,k}^{(t)}}^H  \mathbf{a}(\omega_\mathrm{t}(\theta_g))\mathbf{a}(\omega_\mathrm{t}(\theta_g))^H(\mathbf{w}_{i,k}-{\mathbf{w}_{i,k}^{(t)}})\bigg)\nonumber\\
&\overset{(d)}{=} \sum_{k=1}^K \bigg(
2\Re\{{\mathbf{w}_{i,k}^{(t)}}^H \mathbf{a}(\omega_\mathrm{t}(\theta_g))\mathbf{a}(\omega_\mathrm{t}(\theta_g))^H{\mathbf{w}_{i,k}} \}\bigg)
+c_{i,g}^{(t)},\\
&\overset{(e)}{=}
\Re\{  \mathbf{b}_{i,g}^{(t)H}\mathbf{w}_i \}+c_{i,g}^{(t)}, \label{MM lower bound}
\end{align}
\end{subequations}
\end{small} 

\nid where (a) is obtained by the properties of trace operation and it is further specified according to the definition of $\mathbf{W}_i$ in (b). (c) is obtained according to the inequality between the convex function and its first-order Taylor expansion \cite{MM 2}, in which ${\mathbf{w}_{i,k}^{(t)}}$ is the resulting vector obtained in the $t$-th MM procedure.
(d) and (e) are rearranged with the following definitions for brevity
\begin{subequations}
\begin{align}
\mathbf{b}_{i,g}^{(t)}&\triangleq
2\left[\begin{array}{lccc}
&\hspace{-2ex}\mathbf{a}(\omega_\mathrm{t}(\theta_g))\mathbf{a}(\omega_\mathrm{t}(\theta_g))^H{\mathbf{w}_{i,1}}^{(t)}\\
&\hspace{-2ex}\mathbf{a}(\omega_\mathrm{t}(\theta_g))\mathbf{a}(\omega_\mathrm{t}(\theta_g))^H{\mathbf{w}_{i,2}}^{(t)}\\
&\hspace{-2ex}\ldots\\
&\hspace{-2ex}\mathbf{a}(\omega_\mathrm{t}(\theta_g))\mathbf{a}(\omega_\mathrm{t}(\theta_g))^H{\mathbf{w}_{i,K}}^{(t)}
\end{array} \right] ,   \label{MM b definition}\\
c_{i,g}^{(t)}&\triangleq \sum_k -{\mathbf{w}_{i,k}^{(t)}}^H \mathbf{a}(\omega_\mathrm{t}(\theta_g))\mathbf{a}(\omega_\mathrm{t}(\theta_g))^H {\mathbf{w}_{i,k}^{(t)}}, \label{MM c definition}\\
\mathbf{w}_i&\triangleq\mathrm{vec}(\mathbf{W}_i). \label{vec w definition}
\end{align}
\end{subequations}

Therefore, replacing the non-convex constraint (\ref{SNR constraint}) with its lower-bounded function (\ref{MM lower bound}), the optimization problem in each MM iteration can be formulated as\vspace{-2ex}

\begin{small}
\begin{subequations} \label{MM problem}
\begin{align}
\max\limits_{\mathbf{w}_i,\forall i} &~~R_\mathrm{c} \label{objective f of MM}\\
\text{s.t.} &~~\sum_i \phi_i (\Re\{  \mathbf{b}_{i,g}^{(t)H}\mathbf{w}_i \}+c_{i,g}^{(t)})
\geq \frac{N_\mathrm{sel}\sigma_\mathrm{s}^2\Gamma_\mathrm{0}}{\sigma_\beta^2 \mathrm{PL}(2d_0)} ,~~\forall g, \label{MM SNR constraint}\\
&~~\sum_{i} \|\mathbf{w}_i\|^2 \leq P_0. \label{MM power constraint}
\end{align}
\end{subequations}
\end{small}

\nid Through iteratively solving it, $\mathbf{w}_i$ can converge to a sub-optimal point \cite{MM 2}.
However, the existence of complicated multiple-ratio objective function (\ref{objective f of MM}) makes the solving of iterative problem (\ref{MM problem}) still challenging.
In order to efficiently address this issue, we propose an FP-based method to further transform this problem into a series of tractable sub-problems.

\subsubsection{FP-based Transformation of Objective Function (\ref{objective f of MM})} \label{FP part}
We observe that the fractional terms of the objective function (\ref{objective f of MM}) are contained in the logarithmic functions, which will hinder the subsequent low-complexity algorithm development.
Therefore, we begin by taking these ratio parts out of the logarithms. 
After that, the FP method is applied to tackle the newly constructed non-convex fractional problem.
The detailed transformations are described below.


By leveraging the \textit{Proposition 1} in \cite{FP 1}, problem (\ref{MM problem}) is first equivalently transformed as
\begin{equation} \label{problem of W-MM-FP1}
\begin{aligned}
\max\limits_{\mathbf{w}_i,\mathbf{r}_i,\forall i}  ~&\sum_{i=0}^{N_\mathrm{s}-1}\Bigg( \sum_{k=1}^{K}\log_2(1+r_{i,k})-\sum_{k=1}^{K}r_{i,k}\\
&~~~~~~+\sum_{k=1}^{K} \frac{(1+r_{i,k})|\mathbf{h}_{i,k}^H\mathbf{w}_{i,k}|^2 }{\sum_{j=1}^K|\mathbf{h}_{i,k}^H\mathbf{w}_{i,j}|^2 +\sigma_\mathrm{c}^2 } \Bigg)\\
\text{s.t.} ~~
&~   (\text{\ref{MM SNR constraint}}),~ (\text{\ref{MM power constraint}}),
\end{aligned}
\end{equation}
where $\mathbf{r}_i\triangleq[r_{i,1},\ldots,r_{i,K}]^T \in \mathbb{R}^{K}$, $\forall i$ are the newly introduced auxiliary variables.
In this equivalent problem (\ref{problem of W-MM-FP1}), the fractional terms in the objective function are successfully taken out from the logarithms.
Then, toward the newly constructed non-convex fractional programming (\ref{problem of W-MM-FP1}), it is further transformed into the following equivalent form by applying \textit{Quadratic Transform} proposed in \cite{FP 2}
\begin{equation} \label{problem of W-MM-FP2}
\begin{aligned}
\min\limits_{\mathbf{w}_i,\mathbf{r}_i,\mathbf{t}_i,\forall i}  &~-\sum_{i=0}^{N_\mathrm{s}-1}\Bigg( \sum_{k=1}^{K}\log_2(1+r_{i,k})-\sum_{k=1}^{K}r_{i,k}\\
&~~~~~~+\sum_{k=1}^{K}\bigg(2\sqrt{1+r_{i,k}} \Re\{t_{i,k}^{*}\mathbf{h}_{i,k}^H \mathbf{w}_{i,k} \}\\
&~~~~~~-|t_{i,k}|^2 \big(\sum_{j=1}^{K} |\mathbf{h}_{i,k}^H\mathbf{w}_{i,j}|^2+\sigma_\mathrm{c}^2 \big)\bigg) \Bigg)\\
\text{s.t.} ~~
&~   (\text{\ref{MM SNR constraint}}),~ (\text{\ref{MM power constraint}}),
\end{aligned}
\end{equation}
where $\mathbf{t}_i\triangleq[t_{i,1},\ldots,r_{i,K}]^T\in \mathbb{R}^{K}$, $\forall i$ are auxiliary variables.

It can be seen that the previous complicated problem (\ref{MM problem}) is now transformed into an equivalent form (\ref{problem of W-MM-FP2}), which is convex with respect to $\mathbf{w}_i$, $\mathbf{r}_i$, and $\mathbf{t}_i$, respectively.
In order to efficiently solve this problem (\ref{problem of W-MM-FP2}), we propose to iteratively update the variables $\mathbf{w}_i$, $\mathbf{r}_i$, and $\mathbf{t}_i$ to find the conditionally optimal solution of one variable given others.
Each update is conducted as follows.

The conditionally optimal $\mathbf{r}_i$ and $\mathbf{t}_i$ can be easily obtained by letting the gradient of the objective function of (\ref{problem of W-MM-FP2}) equal to $0$, which are given as follows
\begin{subequations}\label{FP transform}
\begin{align}
r_{i,k}^{\star}&=\frac{|\mathbf{h}_{i,k}^H\mathbf{w}_{i,k}|^2 }{\sum_{j\neq k}|\mathbf{h}_{i,k}^H\mathbf{w}_{i,j}|^2 +\sigma_\mathrm{c}^2 },~~\forall i,~k,\label{FP r update}\\
t_{i,k}^{\star}&= \frac{\sqrt{1+r_{i,k}}\mathbf{h}_{i,k}^H \mathbf{w}_{i,k}   }{ \sum_{j=1}^{K} |\mathbf{h}_{i,k}^H\mathbf{w}_{i,j}|^2+\sigma_\mathrm{c}^2  },~~\forall i,~k. \label{FP t update}
\end{align}
\end{subequations}

The conditionally optimal $\mathbf{w}_i$ can be obtained by solving the problem (\ref{problem of W-MM-FP2}) with $\mathbf{r}_i$ and $\mathbf{t}_i$ being fixed.
Dropping the constant term in the objective function, the problem for updating $\mathbf{w}_i$ is written as
\begin{equation} \label{problem of W-MM-FP3}
\begin{aligned}
\min\limits_{\mathbf{w}_i,\forall i}  &~\sum_{i=0}^{N_\mathrm{s}-1} f_i(\mathbf{w}_i)\\
\text{s.t.}~
&~~   (\text{\ref{MM SNR constraint}}),~ (\text{\ref{MM power constraint}}),
\end{aligned}
\end{equation}
where we define the objective function as
\begin{subequations}
\begin{align}
f_i(\mathbf{w}_i)&\triangleq -\sum_{k=1}^{K}\bigg(2\sqrt{1+r_{i,k}} \Re\{t_{i,k}^{*}\mathbf{h}_{i,k}^H \mathbf{w}_{i,k} \}\\
&~~~~~~~~-|t_{i,k}|^2 \big(\sum_{j=1}^{K} |\mathbf{h}_{i,k}^H\mathbf{w}_{i,j}|^2+\sigma_\mathrm{c}^2 \big)\bigg) \nonumber\\
&=
\mathbf{w}_i^H\mathbf{T}_i\mathbf{w}_i
-2\Re\{\widehat{\mathbf{h}}_{i}^{H}\mathbf{w}_i\},  \label{fi definition}
\end{align}
\end{subequations}
with
\begin{subequations} \label{T i definitions}
\begin{align}
\mathbf{T}_i&\triangleq\mathbf{I}_{K}\otimes(\sum_{j=1}^K|t_{i,j}|^2\mathbf{h}_{i,j}\mathbf{h}_{i,j}^H),\\
\widehat{\mathbf{h}}_{i}&\triangleq[\sqrt{1+r_{i,1}}t_{i,1}\mathbf{h}_{i,1}^T,\ldots,\sqrt{1+r_{i,K}}t_{i,K}\mathbf{h}_{i,K}^T ]^T.
\end{align}
\end{subequations}

After the previous MM-FP transformation, the primal problem (\ref{beamforming problem}) is successfully converted into a series of convex sub-problems (\ref{problem of W-MM-FP3}), which will greatly simplify the optimization and reduce the computational complexity.
However, the dimension of the variable in (\ref{problem of W-MM-FP3}) is too high, which still greatly hinders the efficient beamforming design.
To facilitate the beamforming design with lower computational complexity, we further propose an neADMM based algorithm to decompose the high-dimensional problem (\ref{problem of W-MM-FP3}) into multiple low-dimensional sub-problems and provide closed-form solutions.
In the meantime, this also enables the usage of powerful parallel computing to further accelerate the speed of optimizations.
\begin{figure*}[!b]
    \normalsize
     \hrulefill
    \begin{small}
    \begin{align}
    \mathcal{L}(\mathbf{w}_i,\mathbf{x}_i,p_i,&\Gamma_{i,g},\boldsymbol{\mu}_i,\nu_i,\gamma_{i,g}, \forall g,\forall i)
    =\sum_{i}f_i(\mathbf{w}_i)+\sum_{g,i,\phi_i=1}\mathbb{I}_{\Upsilon_{i,g}}+\sum_{i}\mathbb{I}_{\mathcal{P}_i}
    + \frac{\rho_1}{2}\sum_{i} \left\| \mathbf{x}_i-\mathbf{w}_i  + \frac{\boldsymbol{\mu}_i}{\rho_1} \right\|^2
     - \sum_{i} \frac{\|\boldsymbol{\mu}_i\|^2}{2\rho_1}  \label{AL of neADMM}\\
    &+ \frac{\rho_2}{2}\sum_{i} \left|\| \mathbf{x}_i\|^2-p_i   + \frac{\nu_i}{\rho_2} \right|^2
     - \sum_{i} \frac{\nu_i^2}{2\rho_2}
     +\frac{\rho_3}{2}\sum_{g,i,\phi_i=1}\left|\Re\{  \mathbf{b}_{i,g}^{(t)H}\mathbf{w}_i \}+c_{i,g}^{(t)}-\Gamma_{i,g}+\frac{\gamma_{i,g}}{\rho_3} \right|^2- \sum_{g,i,\phi_i=1} \frac{\gamma_{i,g}^2}{2\rho_3} .\nonumber
    \end{align}
    \end{small}

    \vspace*{4pt}
\end{figure*}
\subsubsection{neADMM Transformation for Problem (\ref{problem of W-MM-FP3})}

The neADMM is an extension of the classic ADMM framework for solving the nonlinear-equality constrained problems, which is recently proposed in \cite{neADMM 1}.
Before utilizing it to decompose the high-dimensional optimization (\ref{problem of W-MM-FP3}) into a series of low-dimensional sub-problems, we should first get the variable into the separable form with equality constraints.
The detailed transformation is conducted in the following.

Observing from the problem (\ref{problem of W-MM-FP3}), each sub-variable $\mathbf{w}_i$, $\forall i$ is already separable with each other in the objective function, but is coupled with each other in the constraints (\ref{MM SNR constraint}) and (\ref{MM power constraint}) due to the sum operation.
Therefore, we propose to introduce auxiliary variables $\Gamma_{i,g} \in \mathbb{R}$ and $p_i \in \mathbb{R}$ to convert problem (\ref{problem of W-MM-FP3}) into the following equivalent optimization
\begin{subequations} \label{problem ADMM1}
\begin{align}
&\min\limits_{\mathbf{w}_i,p_i,\Gamma_{i,g},\forall i,g} \sum_{i=0}^{N_\mathrm{s}-1}f_i(\mathbf{w}_i) \label{obj f of ADMM1}\\
&\qquad~\text{s.t.}
 ~~\Re\{  \mathbf{b}_{i,g}^{(t)H}\mathbf{w}_i \}+c_{i,g}^{(t)}=  \Gamma_{i,g},~\forall g,~i,~\phi_i=1,  \label{SNR constraint i}\\
&\qquad\qquad\|\mathbf{w}_i\|^2 = p_i,~~ \forall i,  \label{power constrant i}\\
&\qquad\qquad\sum_{i,\phi_i=1} \Gamma_{i,g} \geq \frac{N_\mathrm{sel}\sigma_\mathrm{s}^2\Gamma_\mathrm{0}}{\sigma_\beta^2 \mathrm{PL}(2d_0)} ,~~\forall g,  \label{gama constraint}\\
&\qquad\qquad\sum_{i=0}^{N_\mathrm{s}-1} p_i \leq P_0. \label{p constraint}
\end{align}
\end{subequations}
Facilitated by the above transformation, problem (\ref{problem ADMM1}) is already separable on each sub-variable $\mathbf{w}_i$, $\forall i$.
Nevertheless, we notice that each sub-variable $\mathbf{w}_i$ is constrained by multiple linear constraints (\ref{SNR constraint i}) and one spherical constraint (\ref{power constrant i}), which prevent subsequent efficient closed-form solution finding.
Therefore, we further introduce an auxiliary variable $\mathbf{x}_i\in \mathbb{C}^{N_\mathrm{t} K}$ to transform problem (\ref{problem ADMM1}) into the following equivalent form
\begin{subequations} \label{problem ADMM2}
\begin{align}
&\min \limits_{\mathbf{w}_i,\mathbf{x}_i,p_i,\Gamma_{i,g},\forall i,g}  \sum_{i=0}^{N_\mathrm{s}-1}f_i(\mathbf{w}_i) \label{obj f of ADMM2}\\
&\qquad~\text{s.t.} ~~
\Re\{  \mathbf{b}_{i,g}^{(t)H}\mathbf{w}_i \}+c_{i,g}^{(t)}=  \Gamma_{i,g},~\forall g,~i,~\phi_i=1, \label{ne1}\\
&\qquad\qquad~ \mathbf{w}_i=\mathbf{x}_i, ~~\forall i,\label{ne2}\\
&\qquad\qquad~  \|\mathbf{x}_i\|^2 = p_i,~~ \forall i, \label{ne3}\\
&\qquad\qquad~  \sum_{i,\phi_i=1}  \Gamma_{i,g} \geq \frac{N_\mathrm{sel}\sigma_\mathrm{s}^2\Gamma_\mathrm{0}}{\sigma_\beta^2 \mathrm{PL}(2d_0)},~~\forall g, \label{ie1}\\
&\qquad\qquad~ \sum_{i=0}^{N_\mathrm{s}-1} p_i \leq P_0. \label{ie2}
\end{align}
\end{subequations}

To accommodate the neADMM framework, we define the feasible regions of inequality constraints (\ref{ie1}) and (\ref{ie2}) as sets ${\Upsilon}_{i,g}$ and $\mathcal{P}_{i}$, respectively.
The related indicator functions of these sets are
\begin{subequations}
\begin{align}
\mathbb{I}_{\Upsilon_{i,g}}&\triangleq
\begin{cases}
0, &\Gamma_{i,g}\in {\Upsilon_{i,g}},\\
+\infty, &\mathrm{otherwise},
\end{cases} ~~\forall g,~i,~\phi_i=1,  \label{Id1}  \\
\mathbb{I}_{\mathcal{P}_i}&\triangleq
\begin{cases}
0,   &  p_i\in \mathcal{P}_i,\\
+\infty,& \mathrm{otherwise},
\end{cases} ~~\forall i.\label{Id2}
\end{align}
\end{subequations}
Then, by removing inequality constraints (\ref{ie1}) and (\ref{ie2}) and adding the feasibility indicator functions (\ref{Id1}) and (\ref{Id2}) into the objective function (\ref{obj f of ADMM2}), problem (\ref{problem ADMM2}) is thus transformed to
\begin{equation}\label{final form for neADMM}
\begin{aligned}
&\min\limits_{\mathbf{w}_i,\mathbf{x}_i,p_i,\Gamma_{i,g},\forall i,g}~
\sum_{i}f_i(\mathbf{w}_i)+\sum_{g,i,\phi_i=1}\mathbb{I}_{\Upsilon_{i,g}}+\sum_{i}\mathbb{I}_{\mathcal{P}_i} \\
&\qquad~\text{s.t.}
~~\Re\{  \mathbf{b}_{i,g}^{(t)H}\mathbf{w}_i \}+c_{i,g}^{(t)}=  \Gamma_{i,g}, ~~\forall g,~i,~\phi_i=1,\\
&\qquad\qquad~\mathbf{w}_i=\mathbf{x}_i, \forall i,\\
&\qquad\qquad~\| \mathbf{x}_i\|^2= p_i, \forall i.
\end{aligned}
\end{equation}

Finally, the previous problem (\ref{problem of W-MM-FP3}) is transformed into the equivalent form (\ref{final form for neADMM}), whose solution can be obtained by optimizing its augmented Lagrangian (AL) function in an neADMM manner \cite{neADMM 1}.
Specifically, the AL function of problem (\ref{final form for neADMM}) is expressed as (\ref{AL of neADMM}), shown at the bottom of this page, where $\rho_1$, $\rho_2$, and $\rho_3$ are positive penalty parameters, $\boldsymbol{\mu}_i\in \mathbb{C}^{N_\mathrm{t}K}$, $\nu_i\in \mathbb{C}$, and $\gamma_{i,g}\in \mathbb{C}$ are dual variables.
Through alternately updating the variables $\mathbf{w}_i$, $\mathbf{x}_i$, $\Gamma_{i,g}$, $p_i$, $\boldsymbol{\mu}_i$, $\nu_i$, and $\gamma_{i,g}$ by minimizing AL (\ref{AL of neADMM}), problem (\ref{final form for neADMM}) can be solved. In the following, closed-form solutions for each update are provided in detail.

\textit{\textbf{neADMM Update of $\mathbf{w}_i$}:} With given $\mathbf{x}_i$, $\Gamma_{i,g}$, $p_i$, $\boldsymbol{\mu}_i$, $\nu_i$, and $\gamma_{i,g}$, the optimization for updating $\mathbf{w}_i$ is formulated as
\begin{align}
&\min\limits_{\mathbf{w}_i}\label{neADMM update of W}
~f_{\mathrm{w},i}^{(t)}(\mathbf{w}_i)\triangleq f_i(\mathbf{w}_i) + \frac{\rho_1}{2} \left\| \mathbf{x}_i-\mathbf{w}_i  + \frac{\boldsymbol{\mu}_i}{\rho_1} \right\|^2\\
&~~~~~~~~+\frac{\rho_3}{2}\phi_i\sum_{g}\left|\Re\{  \mathbf{b}_{i,g}^{(t)H}\mathbf{w}_i \}+c_{i,g}^{(t)}-\Gamma_{i,g}+\frac{\gamma_{i,g}}{\rho_3} \right|^2. \nonumber
\end{align}
In order to efficiently deal with this complex-valued problem, we further transform it into a concise real-valued problem by defining\vspace{-2ex}

\begin{small}
\begin{subequations}\label{U1 U2 A b definitions}
\begin{align}
\widehat{\mathbf{w}}_i\triangleq&
\left[\begin{array}{lccc}
&\hspace{-2ex}\Re\{\mathbf{w}_i\}\\
&\hspace{-2ex}\Im\{\mathbf{w}_i\}
\end{array} \right],\\
\mathbf{U}\triangleq&[\mathbf{I}_{N_\mathrm{t}K},\jmath\mathbf{I}_{N_\mathrm{t}K} ], \label{U1}\\
\mathbf{A}_{i}^{(t)}\triangleq &\mathbf{U}^H (\mathbf{T}_i+\frac{\rho_1}{2}\mathbf{I}_{N_\mathrm{t}K})\mathbf{U}\\
 &+ \frac{\rho_3}{2}\phi_i\sum_{g}
 \left[\begin{array}{lccc}
&\hspace{-2.5ex}\Re\{\mathbf{b}_{i,g}^{(t)}\}\\
&\hspace{-2.5ex}\Im\{\mathbf{b}_{i,g}^{(t)}\}
\end{array} \right]
\left[\Re\{\mathbf{b}_{i,g}^{(t)}\}^T,\Im\{\mathbf{b}_{i,g}^{(t)}\}^T\right],
\label{A definition}\\
\mathbf{b}_i^{(t)}\! \triangleq &
\left[\begin{array}{lccc}
&\hspace{-3ex}\Re\{2\widehat{\mathbf{h}}_{i}+( \mathbf{x}_i\hspace{-0.5ex}+\hspace{-0.5ex} \frac{\boldsymbol{\mu}_i}{\rho_1})\hspace{-0.5ex}-\hspace{-0.5ex}\phi_i\sum_{g}(c_{i,g}^{(t)}-\Gamma_{i,g}\hspace{-0.5ex}+\hspace{-0.5ex}\frac{\gamma_{i,g}}{\rho_3})\mathbf{b}_{i,g}^{(t)}\}\\
&\hspace{-3ex}\Im\{2\widehat{\mathbf{h}}_{i}+( \mathbf{x}_i\hspace{-0.5ex}+\hspace{-0.5ex} \frac{\boldsymbol{\mu}_i}{\rho_1})\hspace{-0.5ex}-\hspace{-0.5ex}\phi_i\sum_{g}(c_{i,g}^{(t)}-\Gamma_{i,g}\hspace{-0.5ex}+\hspace{-0.5ex}\frac{\gamma_{i,g}}{\rho_3})\mathbf{b}_{i,g}^{(t)}\}
\end{array} \hspace{-1ex}\right].
\label{b definition}
\end{align}
\end{subequations}
\end{small}\vspace{-2.5ex}

\nid Then, the equivalent compact real-valued form of problem (\ref{neADMM update of W}) is expressed as
\begin{align}
&\min\limits_{\widehat{\mathbf{w}}_i}\label{neADMM update of W mao}
~f_{\mathrm{w},i}^{(t)}(\widehat{\mathbf{w}}_i)=\widehat{\mathbf{w}}_i^T \mathbf{A}_{i}^{(t)} \widehat{\mathbf{w}}_i- \mathbf{b}_i^{(t)T}\widehat{\mathbf{w}}_i.
\end{align}
This is an unconstrained convex optimization, whose optimal solution is the point satisfying
\begin{equation}
\begin{aligned}
\frac{\partial f_{\mathrm{w},i}^{(t)}(\widehat{\mathbf{w}}_i)}{ \partial \widehat{\mathbf{w}}_i^*}
=& (\mathbf{A}_i^{(t)}+{\mathbf{A}_i^{(t)}}^T) \widehat{\mathbf{w}}_i- \mathbf{b}_i^{(t)}=\mathbf{0}.\label{linear equation of W}
\end{aligned}
\end{equation}
This equation can be easily solved by multiplying both sides of it with Moore-Penrose pseudoinverse $(\mathbf{A}_i^{(t)}+{\mathbf{A}_i^{(t)}}^T)^{\dag}$ when it is solvable.
However, since matrix $(\mathbf{A}_i^{(t)}+{\mathbf{A}_i^{(t)}}^T)$ is semidefinite, equation (\ref{linear equation of W}) may be insolvable, which indicates that the minimization problem (\ref{neADMM update of W mao}) is unbounded below according to the Example 4.5 in \cite{book: Boyd}.
Based on the proof in Appendix A, we have the following proposition.

\nid \textit{Proposition 1:} The minimization problem (\ref{neADMM update of W mao}) is bounded below, which means that equation (\ref{linear equation of W}) is solvable and one of its optimal solutions is
\be \label{optimal w}
\widehat{\mathbf{w}}_i^{\star}= (\mathbf{A}_i^{(t)}+{\mathbf{A}_i^{(t)}}^T)^{\dag}\mathbf{b}_i^{(t)}.
\ee
\textit{Proof:} See Appendix A.

\textit{\textbf{neADMM Update of $\mathbf{x}_i$}:} With given $\mathbf{w}_i$, $\Gamma_{i,g}$, $p_i$, $\boldsymbol{\mu}_i$, $\nu_i$, and $\gamma_{i,g}$, the optimization for updating $\mathbf{x}_i$ is formulated as
\begin{equation}\small\label{neADMM update of X}
\begin{aligned}
\min\limits_{\mathbf{x}_i}~
h(\mathbf{x}_i)\triangleq
\!\frac{\rho_1}{2}\!\left\| \mathbf{x}_i\!-\!\mathbf{w}_i \!+\! \frac{\boldsymbol{\mu}_i}{\rho_1} \right\|^2 \!\!\!
+\!\frac{\rho_2}{2}\!\left|\| \mathbf{x}_i\|^2\!\!-\!p_i \!+\! \frac{\nu_i}{\rho_2} \right|^2.
\end{aligned}
\end{equation}
Although this optimization is non-convex, it has been proved in \cite{neADMM 1} that the necessary condition for $\mathbf{x}_i$ to be a minimizer of (\ref{neADMM update of X}) is
\begin{equation}\small \label{equation of x}
\begin{aligned}
\frac{\partial h(\mathbf{x}_i)}{\partial \mathbf{x}_{i}^* }\!=\!\frac{\rho_1}{2}(\mathbf{x}_i-\!(\mathbf{w}_i\!-\!\frac{\boldsymbol{\mu}_i}{\rho_1})) +{\rho_2} (\|\mathbf{x}_i\|^2\hspace{-1ex}-\!p_i   \!+\! \frac{\nu_i}{\rho_2} )\mathbf{x}_i=\mathbf{0}.
\end{aligned}
\end{equation}
By finding out all the solutions of (\ref{equation of x}) and picking out the one which has a minimum value of $h(\mathbf{x}_i)$, we can obtain the optimal solution of (\ref{neADMM update of X}). The detailed solving process is presented below.

Firstly, we denote the solutions of equation (\ref{equation of x}) as $\mathbf{x}_i^{\mathrm{s}}$.
After rearranging of equation (\ref{equation of x}), solution $\mathbf{x}_i^{\mathrm{s}}$ is satisfying
\begin{equation}\label{theorem equation1}
\begin{aligned}
\mathbf{x}_i^{\mathrm{s}}=
\frac{ \mathbf{w}_i-\frac{\boldsymbol{\mu}_i}{\rho_1}  }
{1+\frac{2\rho_2}{\rho_1} \|\mathbf{x}_i^{\mathrm{s}}\|^2-\frac{2\rho_2}{\rho_1}p_i   + \frac{2\nu_i}{\rho_1} }.
\end{aligned}
\end{equation}
Now, the only issue for obtaining $\mathbf{x}_i^{\mathrm{s}}$ is to have its norm $\|\mathbf{x}_i^{\mathrm{s}}\|$, which appears on the right side of equation (\ref{theorem equation1}).
Toward this, we take the norm on both sides of the equation (\ref{theorem equation1}) as follows
\begin{equation} \label{proof equation}
\begin{aligned}
\|\mathbf{x}_i^{\mathrm{s}}\|=
\frac{ \|\mathbf{w}_i-\frac{\boldsymbol{\mu}_i}{\rho_1} \| }
{|1+\frac{2\rho_2}{\rho_1} \|\mathbf{x}_i^{\mathrm{s}}\|^2-\frac{2\rho_2}{\rho_1}p_i   + \frac{2\nu_i}{\rho_1} |}.
\end{aligned}
\end{equation}
This is a typical cubic equation of $\|\mathbf{x}_i^{\mathrm{s}}\|$ and can be easily solved by the general cubic formula \cite{cubic formula} with closed-form solutions.
It should be noted that $\|\mathbf{x}_i^{\mathrm{s}}\|$ of cubic equation (\ref{proof equation}) possibly has up to $6$ different values, correspondingly, (\ref{theorem equation1}) will have up to $6$ possible values.
Thus, the following step for obtaining the optimal solution to the problem (\ref{neADMM update of X}) is
\begin{equation} \label{optimal x}
\begin{aligned}
\mathbf{x}_i^{\star} \triangleq\min\limits_{\mathbf{x}_i^{\mathrm{s}}}~~
h(\mathbf{x}_i^{\mathrm{s}}).
\end{aligned}
\end{equation}

\textit{\textbf{neADMM Update of $\Gamma_{i,g}$}:} With given $\mathbf{w}_i$, $\mathbf{x}_i$, $p_i$, $\boldsymbol{\mu}_i$, $\nu_i$, and $\gamma_{i,g}$, the optimization for updating $\Gamma_{i,g}$ is formulated as
\be\small
\min\limits_{\Gamma_{i,g},\forall i,\phi_i=1} \sum_{i,\phi_i=1}\mathbb{I}_{\Upsilon_{i,g}}\hspace{-1ex}+ \hspace{-1ex}\sum_{i,\phi_i=1}
\left|\Re\{  \mathbf{b}_{i,g}^{(t)H}\mathbf{w}_i \}+c_{i,g}^{(t)}+\frac{\gamma_{i,g}}{\rho_3}-\Gamma_{i,g} \right|^2.
\ee
According to the definition of indicator $\mathbb{I}_{\Upsilon_{i,g}}$ in (\ref{Id1}), this problem can be equivalently transformed into a convex quadratic problem:\vspace{-2ex}

\begin{small}
\begin{subequations} \label{neADMM update of Gama}
\begin{align}
\min\limits_{\Gamma_{i,g},\forall i,\phi_i=1}&f_\Gamma^{(t)}(\Gamma_{i,g},\forall i,\phi_i=1) \label{obj f of update of Gama}\\
&\triangleq\sum_{i,\phi_i=1}
\left|\Re\{  \mathbf{b}_{i,g}^{(t)H}\mathbf{w}_i \}+c_{i,g}^{(t)}+\frac{\gamma_{i,g}}{\rho_3}-\Gamma_{i,g} \right|^2  \nonumber \\
\text{s.t.} ~~
&\sum_{i,\phi_i=1} \Gamma_{i,g} \geq  \frac{N_\mathrm{sel}\sigma_\mathrm{s}^2\Gamma_\mathrm{0}}{\sigma_\beta^2 \mathrm{PL}(2d_0)}. \label{constraint of update of Gama}
\end{align}
\end{subequations}
\end{small}

\nid Problem (\ref{constraint of update of Gama}) is quadratic programming, which can be solved by some existing algorithms, e.g., interior-point method and active set method \cite{book: active set}.
We see that each of the variables $\Gamma_{i,g}$, $\forall i,\phi_i=1$ in the objective function (\ref{obj f of update of Gama}) is separate from each other, which allows problem (\ref{neADMM update of Gama}) to be more easily handled.
Therefore, in an effort to further alleviate the computing burden, we provide a closed-form solution by exploiting this characteristic as follows.

By setting the partial derivation with respect to $\Gamma_{i,g}$ to zero
\begin{align} 
g_\Gamma(\Gamma_{i,g})&\triangleq\frac{\partial f_\Gamma^{(t)}(\Gamma_{i,g},\forall i,\phi_i=1)}{\partial \Gamma_{i,g}} \label{partial g of Gama} \\
 &=\Gamma_{i,g}- \Re\{  \mathbf{b}_{i,g}^{(t)H}\mathbf{w}_i \}+c_{i,g}^{(t)}+\frac{\gamma_{i,g}}{\rho_3}=0,\nonumber
\end{align}
the unconditional optimal ${\widehat{\Gamma}_{i,g}}^{\star}$ can be calculated as
\be
{\widehat{\Gamma}_{i,g}}^{\star}= \Re\{  \mathbf{b}_{i,g}^{(t)H}\mathbf{w}_i \}+c_{i,g}^{(t)}+\frac{\gamma_{i,g}}{\rho_3},~~ \forall i,~\phi_i=1.
\ee
Given a variable $\Delta\Gamma \in \mathbb{R}$ of any value and substituting ${\widehat{\Gamma}_{i,g}}^{\star}+\Delta\Gamma$ into the gradient expression (\ref{partial g of Gama}), we can have
\be
g_\Gamma({\widehat{\Gamma}_{i,g}}^{\star}+\Delta\Gamma)= \Delta\Gamma, ~~\forall i,~ \phi_i=1.
\ee
This gradient indicates that when the deviation from the unconditional optimal point ${\widehat{\Gamma}_{i,g}}^{\star}$, $\forall i,\phi_i=1$ is the same, the associated increment caused to the objective function (\ref{obj f of update of Gama}) is the same.
Moreover, as the deviation value increases linearly, the objective function will grow faster and faster.
Therefore, with considering constraint (\ref{constraint of update of Gama}), the optimal solution of problem (\ref{neADMM update of Gama}) can be calculated as
\be\label{optimal Gama}\small
\Gamma_{i,g}^{\star}\triangleq {\widehat{\Gamma}_{i,g}}^{\star}\!+\! \max\left\{(\frac{N_\mathrm{sel}\sigma_\mathrm{s}^2\Gamma_\mathrm{0}}{\sigma_\beta^2 \mathrm{PL}(2d_0)}-\sum_{i,\phi_i=1} {\widehat{\Gamma}_{i,g}}^{\star})/N_\mathrm{sel} ,0\right\}.
\ee

\textit{\textbf{neADMM Update of $p_i$}:} With given $\mathbf{w}_i$, $\mathbf{x}_i$, $\Gamma_{i,g}$, $\boldsymbol{\mu}_i$, $\nu_i$, and $\gamma_{i,g}$, the optimization for updating $p_i$ is formulated as
\begin{equation}
\begin{aligned}
\min\limits_{p_i,\forall i} ~\sum_{i}\mathbb{I}_{\mathcal{P}_i}+\sum_i\left|\| \mathbf{x}_i\|^2-p_i   + \frac{\nu_i}{\rho_2} \right|^2.
\end{aligned}
\end{equation}
According to the definition of indicator (\ref{Id2}), this problem can be equivalently transformed into the following form
\begin{equation} \label{neADMM update of p}
\begin{aligned}
\min\limits_{p_i,\forall i} ~&\sum_i\left|\| \mathbf{x}_i\|^2-p_i + \frac{\nu_i}{\rho_2} \right|^2\\
\text{s.t.} ~~
&\sum_i p_i\leq P_0.
\end{aligned}
\end{equation}
Problem (\ref{neADMM update of p}) has the same property as that of the problem (\ref{neADMM update of Gama}), so it can be solved efficiently with a similar solving strategy for the problem (\ref{neADMM update of Gama}).
Due to the space limitation, the derivations are omitted here, and the optimal solution of (\ref{neADMM update of p}) is
\be \label{optimal p}
p_i^{\star}={\widehat{p}_i}^{\star}- \max\left\{ (\sum_i {\widehat{p}_i}^{\star}-P_0)/N_\mathrm{s} ,0\right\},
\ee
with unconditional optimal ${\widehat{p}_i}^{\star}$ as
\be
{\widehat{p}_i}^{\star}=\| \mathbf{x}_i\|^2 + \frac{\nu_i}{\rho_2}.
\ee

\textit{\textbf{neADMM Update of $\boldsymbol{\mu}_i$, $\nu_i$, and $\gamma_{i,g}$}:} After obtaining $\mathbf{w}_i$, $\mathbf{x}_i$, $\Gamma_{i,g}$, and $p_i$, the dual variables $\boldsymbol{\mu}_i$, $\nu_i$, and $\gamma_{i,g}$ are updated by
\begin{subequations} \label{neADMM update of mu and nu}
\begin{align}
\boldsymbol{\mu}_i^\star&:= \boldsymbol{\mu}_i+ \rho_1 (\mathbf{x}_i-\mathbf{w}_i ),\\
\nu_i^\star&:= \nu_i+ \rho_2 (\| \mathbf{x}_i\|^2-p_i),\\
\gamma_{i,g}^\star&:= \gamma_{i,g}+ \rho_3 \big(\Re\{  \mathbf{b}_{i,g}^{(t)H}\mathbf{w}_i \}+c_{i,g}^{(t)}-\Gamma_{i,g}\big).
\end{align}
\end{subequations}

\begin{algorithm}[t]\begin{small}
  \caption{Proposed MM-FP-neADMM based Algorithm for Beamforming Design (\ref{beamforming problem})}
  \label{Algorithm Pseudocode}
  \begin{algorithmic}[1]
    \REQUIRE $\mathbf{h}_{i,k}$, $\forall i,k$, $\sigma_\mathrm{c}$, $\sigma_\mathrm{s}$, $\sigma_\mathrm{b}$, $d_0$, $\theta_g$, $\forall g$, $\Gamma_\mathrm{0}$, $P_0$, $\rho_1$, $\rho_2$, $\rho_3$.
    \ENSURE  $\mathbf{W}_{i}^{\star}$, $\forall i$.
    \STATE {Initialize $\mathbf{W}_{i}$ by (\ref{W initialization}), $\mathbf{X}_{i}=\mathbf{W}_{i}$, $p_i=\|\mathbf{W}_i\|^2$, $\Gamma_{i,g}=\mathbf{a}\left(\omega_\mathrm{t}(\theta_g)\right)^H \mathbf{W}_i \mathbf{W}_i^H\mathbf{a}\left(\omega_\mathrm{t}(\theta_g)\right)$, $\boldsymbol{\mu}_i=\mathbf{0}$, $\nu_i={0}$, $\gamma_{i,g}=0$, $\forall i,g$, $t = 0$.}
    \REPEAT
    \STATE {Obtain $\mathbf{b}_{i,g}^{(t)}$ and $c_{i,g}^{(t)}$ by (\ref{MM b definition}) and (\ref{MM c definition}), respectively;}
    \STATE {Obtain $r_{i,k}^{\star}$ and $t_{i,k}^{\star}$ by equations (\ref{FP r update}) and (\ref{FP t update}), respectively;}
    \STATE {Update $\mathbf{w}_i$ by equation (\ref{optimal w});}
    \STATE {Update $\mathbf{x}_{i}$: Firstly, obtain all the roots $\|\mathbf{x}_i^{\mathrm{s}}\|$ of cubic equation (\ref{proof equation}). Then, calculate corresponding stationary points $\mathbf{x}_i^\mathrm{s}$ based on (\ref{theorem equation1}) with $\|\mathbf{x}_i^{\mathrm{s}}\|$. Finally, select out $\mathbf{x}_{i}$ by (\ref{optimal x});}
    \STATE {Update $\Gamma_{i,g}$ by equation (\ref{optimal Gama});}
    \STATE {Update $p_{i}$ by equation (\ref{optimal p});}
    \STATE {Update $\boldsymbol{\mu}_i$, $\nu_i$, and $\gamma_{i,g}$ by (\ref{neADMM update of mu and nu});}
    \STATE {$t:= t + 1$;}
    \UNTIL {Convergence}.
    \STATE {Return $\mathbf{W}_{i}^{\star}=\mathbf{W}_{i}$, $\forall i$.}
  \end{algorithmic}\end{small}
\end{algorithm} 

\subsection{Summary, Initialization, and Complexity Analysis}

\begin{figure*}[!b]
    \normalsize
     \hrulefill
    \begin{small}
    \begin{subequations} \label{neADMM update of residual}
    \begin{align}
    r\triangleq &
    \sqrt{\sum_i \Big(\|\mathbf{x}_i-\mathbf{w}_i\|^2
    + \left| \| \mathbf{x}_i\|^2-p_i \right|^2 \Big)
    +\sum_{g,i,\phi_i=1}\Big|\Re\{  \mathbf{b}_{i,g}^{(t)H}\mathbf{w}_i \}+c_{i,g}^{(t)}-\Gamma_{i,g} \Big|^2}  \label{primal residual norm}\\
    s\triangleq & \sqrt{\sum_i\Big(\| \rho_1(\mathbf{w}_i^{(t)}-\mathbf{w}_i^{(t-1)})  \|^2+| \rho_2(p_i^{(t)}-p_i^{(t-1)})|^2\Big)
    + \sum_{g,i,\phi_i=1} \| \rho_3\mathbf{b}_{i,g}^{(t)}(\Gamma_{i,g}^{(t)}-\Gamma_{i,g}^{(t-1)}) \|^2}\label{dual residual norm}
    \end{align}
    \end{subequations}
    \end{small}

    \vspace*{4pt}
\end{figure*}

Based on the above development, the proposed algorithm for the beamforming design (\ref{beamforming problem}) is straightforward and summarized in Algorithm \ref{Algorithm Pseudocode}.
In summary, the beamforming matrices $\mathbf{W}_i$, $\forall i$, are obtained by iteratively minimizing the AL (\ref{AL of neADMM}) of the final transformed problem (\ref{final form for neADMM}) in an MM-FP-neADMM manner until convergence.
Specifically, in each iteration, we first update the MM-based surrogate constraint (\ref{ne1}) determined by $\mathbf{b}_{i,g}^{(t)}$ and $c_{i,g}^{(t)}$ via (\ref{MM b definition}) and (\ref{MM c definition}), respectively.
Then update the FP-based objective function (\ref{obj f of ADMM2}) determined by $r_{i,k}^{\star}$ and $t_{i,k}^{\star}$ via (\ref{FP r update}) and (\ref{FP t update}), respectively.
Afterward, the updated AL (\ref{AL of neADMM}) is obtained by the neADMM procedure, in which $\mathbf{w}_i$, $\mathbf{x}_{i}$, $\Gamma_{i,g}$, $p_{i}$, and dual variables ($\boldsymbol{\mu}_i$, $\nu_i$, $\gamma_{i,g}$) are updated via (\ref{optimal w}), (\ref{optimal x}), (\ref{optimal Gama}), (\ref{optimal p}), and (\ref{neADMM update of mu and nu}), respectively.

Since the starting point will affect the convergence speed and result of the proposed iterative algorithm, we propose an initialization method to ensure a good convergence in the following.
Considering that the beamforming matrix $\mathbf{W}_{i}$ in (\ref{beamforming problem}) is optimized for maximizing communication service restricted by sensing and power constraint.
Therefore, we propose one effective initialization method that focuses the beamforming toward the interested sensing region to provide a feasible starting point.
Specifically, $\mathbf{W}_i$ is initialized as follows
\begin{equation} \small \label{W initialization}
\begin{aligned}
\mathbf{w}_{i,k}=\sqrt{\frac{P_0}{N_\mathrm{s}N_\mathrm{t}K}}\mathbf{a}\big(\omega_\mathrm{t}(\theta_\mathrm{a}\!+\!(k-1)\frac{ \theta_\mathrm{b}-\theta_\mathrm{a}}{K-1})\big),
\end{aligned}
\end{equation}
which is designed to be aligned with $K$ line-of-sight (LoS) channels uniformly distributed in the interested angular range $\theta_\mathrm{a} \sim \theta_\mathrm{b}$, respectively. $\sqrt{\frac{P_0}{N_\mathrm{s}N_\mathrm{t}K}}$ is the power scaling coefficient to meet the power budget.

Next, we provide a brief complexity analysis of the proposed Algorithm \ref{Algorithm Pseudocode}. In each iteration, the computational complexity to update all the $\mathbf{b}_{i,g}^{(t)}$ and $c_{i,g}^{(t)}$, $\forall i,g$, is of order $\mathcal{O}(N_\mathrm{s}GN_\mathrm{t}K)$, the computational complexity to update all the $r_{i,k}^{\star}$ and $t_{i,k}^{\star}$, $\forall i,k$, is of order $\mathcal{O}(N_\mathrm{s}N_\mathrm{t}K^2)$, the computational complexity to update all the $\mathbf{w}_i$, $\forall i$, is of order $\mathcal{O}(N_\mathrm{s}(N_\mathrm{t}K)^3)$, and the computational complexity to update all the $\Gamma_{i,g}$, $\forall g,i,\phi_i=1$, is of order $\mathcal{O}(N_\mathrm{s}GN_\mathrm{t}K)$.
Moreover, the left update of $\mathbf{x}_{i}$, $p_{i}$, dual variables ($\boldsymbol{\mu}_i$, $\nu_i$, and $\gamma_{i,g}$), $\forall i$, requires the same order of computational complexity
$\mathcal{O}(N_\mathrm{s}N_\mathrm{t}K)$.
Thus, the computational cost of Algorithm \ref{Algorithm Pseudocode} for beamforming design (\ref{beamforming problem}) mainly depends on the update of $\mathbf{w}_i$, $\forall i$, which is further dominated by the number of transmit antennas $N_\mathrm{t}$ and communication users $K$.
In addition to the main factor $N_\mathrm{t}$ and $K$, the number of subcarriers $N_\mathrm{s}$ and discrete grid number $G$ will also affect the computational cost in a linear relation.


\section{Simulation Results} \label{Simulation part}

\begin{table}[!t]
\centering
\begin{small}
\caption{System settings.}
\begin{tabular}{p{4.3cm} p{1.4cm} p{1.7cm} }
\toprule\label{tab: parameters}
Parameter  &Symbol & Value\\
\hline
Carrier frequency       & $f_\mathrm{c}$    &$28$GHz  \\
Subcarrier spacing      &$\Delta f$         & $120$kHz   \\
Number of subcarriers   &$N_\mathrm{s}$     &$256$ \\
OFDM symbol duration    & $T_\mathrm{d}$    & $8.33\mu \mathrm{s}$    \\
CP duration             & $T_\mathrm{cp}$  & $0. 59\mu \mathrm{s}$  \\
Total symbol duration   &$T$                & $8.92 \mu \mathrm{s}$ \\
Number of OFDM symbols  &$L$                &$128$\\
Number of transmit antennas  &$N_\mathrm{t}$                &$24$\\
Number of receive antennas  &$N_\mathrm{r}$                &$24$\\
Transmit antenna spacing &$d_\mathrm{t}$    &$0.5 c/f_\mathrm{c}$\\
Receive antenna spacing &$d_\mathrm{t}$     &$0.5 c/f_\mathrm{c}$\\
Reference distance      &$d_\mathrm{ref}$   &$1\mathrm{m}$\\
Loss of the reference distance&$c_\mathrm{ref}$ &$-30\mathrm{dB}$\\
Path loss exponent      &$\alpha$           &$2.6$\\
Reflection coefficient power&$\sigma_\beta^2$ &$0 \mathrm{dB}$\\
Communication noise power& $\sigma_\mathrm{c}^2$&$-60\mathrm{dBm}$\\
Sensing noise power     & $\sigma_\mathrm{s}^2$&$-60\mathrm{dBm}$\\
QAM order               & $\Omega$          &$16$\\
DFT point in spatial dimension &$N_\mathrm{a}$  & $N_\mathrm{r}$\\
DFT point in delay dimension &$N_\mathrm{d}$  & $N_\mathrm{s}$\\
DFT point in Doppler dimension &$N_\mathrm{v}$  & $L$\\
Power budget &$P_0$  & $10\mathrm{W}$\\
\bottomrule
\end{tabular}\vspace{-2ex}
\end{small}
\end{table}

In this section, we provide simulation experiments to illustrate the effectiveness of the proposed algorithm and the advantages of the proposed sparsity exploitation strategy for MIMO-OFDM ISAC systems.
Unless otherwise specified, the system parameters are set in Table \ref{tab: parameters}, based on the 3GPP 5G NR high-frequency standard \cite{3GPP 211}.
We consider a scenario with $K=5$ communication users in the following simulations.
The region of interest for sensing is located within a specific angular range $\theta_\mathrm{a}=-10^\circ\sim \theta_\mathrm{b}=10^\circ$ and a specific distance range $0\sim d_0=75\mathrm{m}$. 
To ensure the estimation performance of the interested region within this angular direction $\theta_\mathrm{a}\sim \theta_\mathrm{b}$, the number of grids is set as $G=10$.
Additionally, the penalty parameters are set as $\rho_1=500$, $\rho_2=500$, and $\rho_3=50$.
The communication channel is specifically modeled to incorporate the channel characteristics of the high-frequency band. It takes into account a LoS path and $L_\mathrm{path} = 5$ primary scattering paths from the BS to the user:
\begin{align}
\mathbf{h}_{i,k} \triangleq & \delta_k \sqrt{\mathrm{PL}(d_k)} \mathbf{a}\left(
\frac{2\pi  d_\mathrm{t} \sin\theta_k(f_\mathrm{c}+i\Delta f)}{c}\right) \\
&+ \sum_{l=1}^{L_\mathrm{path}} \delta_{k,l} \sqrt{\mathrm{PL}(d_{k,l})} \mathbf{a}\left(\frac{2\pi  d_\mathrm{t} \sin\theta_{k,l}(f_\mathrm{c}+i\Delta f)}{c}\right), \nonumber
\end{align}
with the first and second terms corresponding to the LoS and scattering paths, respectively.
$d_{k}$ and $\theta_{k}$ denote the length and the DoA of the LoS path, respectively.
$\delta_k$ and $\delta_{k,l}$ denote the receiving coefficient.
$d_{k,l}$ and $\theta_{k,l}$ denote the length and the DoA of the $l$-th scattering path, respectively.
In the simulations, corresponding parameters $\delta_k$, $\delta_{k,l}$, $L_\mathrm{path}$, $d_k$, $d_{k,l}$, $\theta_k$, and $\theta_{k,l}$ are set as $1$, $\delta_l\sim \mathcal{CN}(0.1, 0.01)$, $5$, $d_k\sim \mathcal{U} (50\mathrm{m},100\mathrm{m})$, $d_{k,l}\sim \mathcal{U}(d_k-15\mathrm{m}, d_k+15\mathrm{m})$, $\theta_k\sim \mathcal{U} (-90^{\circ},+90^{\circ})$, and $\theta_l\sim \mathcal{U}(\theta_k -10^{\circ},\theta_k+10^{\circ})$, respectively.

In the following simulations, the proposed sparsity exploitation strategy utilizing the CS-aided estimation method in Sec. \ref{sec: CS-estimation} and the associated beamforming design in Sec. \ref{sec: beamforming} is denoted as ``Proposed, Alg\ref{Algorithm Pseudocode}'', in which the number of selected subcarriers for sensing is set as $N_\mathrm{sel}=N_\mathrm{s}/4=64$.
To illustrate the advantages brought by our proposed sparsity exploitation strategy, we include the algorithms without sparsity exploitation which employs our previous estimation in \cite{our estimation work} and the beamforming design in Sec. \ref{sec: beamforming} with utilizing all the subcarriers for sensing, i.e., $N_\mathrm{sel}=N_\mathrm{s}$, and it is denoted as ``W/o SE, Alg\ref{Algorithm Pseudocode}''. The communication-only design, which employs our previous estimation in \cite{our estimation work} and the beamforming design in Sec. \ref{sec: beamforming} without considering sensing requirement (i.e., $N_\mathrm{sel}=0$), is also included and referred to as ``Commum-only, Alg\ref{Algorithm Pseudocode}''.
To facilitate a comprehensive analysis of the benefits associated with our proposed beamforming algorithm, the state-of-the-art iteratively weighted sum mean-square error minimization (WMMSE) algorithm  \cite{WMMSE} for maximizing communication sum-rate with power constraint is chosen as a benchmark, denoted as ``Commum-only, \cite{WMMSE}''.
Furthermore, a simple heuristic algorithm for solving beamforming design (\ref{beamforming problem}) is introduced for comparison purposes.
Specifically, this heuristic algorithm is conducted in two following steps.
In step one, the beamforming matrix is first initialized in equation (\ref{W initialization}) and then scaled to satisfy the constraint (\ref{SNR constraint}).
The remaining power budget is for maximizing the communication sum-rate with the WMMSE approach in \cite{WMMSE} and then added to the previous beamforming matrix.
The heuristic design with/without sparsity exploitation is denoted as ``W/ SE, Heuristic''/``W/o SE, Heuristic'', respectively.

\begin{figure}[!t]
\centering
\includegraphics[width = 3.5 in]{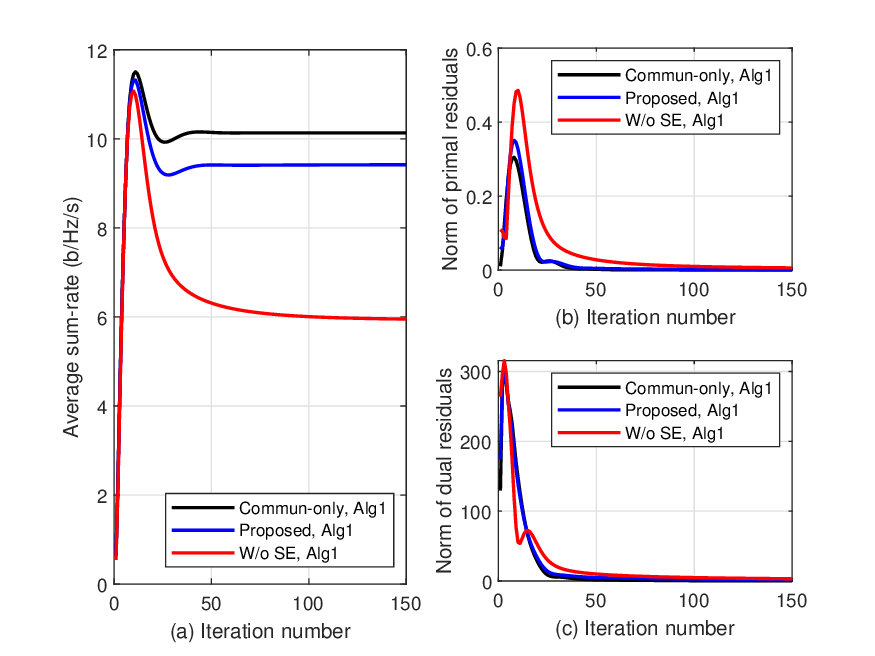}\vspace{-1ex}
\caption{Average sum-rate, the norm of the primal/dual residuals versus the number of iterations ($\Gamma_0=-5\mathrm{dB}$, $P_0=10\mathrm{W}$).} \vspace{-3.5 ex}
\label{iteration fig}
\end{figure}

\begin{figure}[htbp]
\centering
\includegraphics[width = 3.5 in]{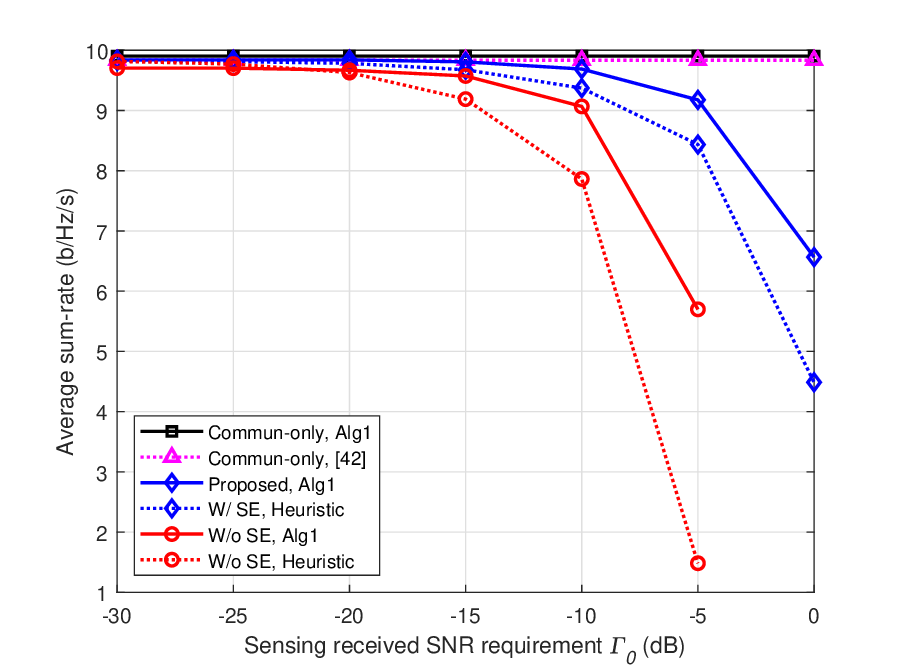}\vspace{-1ex}
\caption{Average sum-rate versus sensing received SNR requirement $\Gamma_0$ ($P_0=10\mathrm{W}$).} \vspace{-2 ex}
 \label{rate SNR fig}
\end{figure}

We start with presenting the convergence performance of the proposed algorithm by plotting the average sum-rate, the norm of the primal and dual residual versus the number of iterations in Fig. \ref{iteration fig}, where the sensing SNR threshold is set as $\Gamma_\mathrm{0}=-5\mathrm{dB}$, the total transmit power is $P_0=10\mathrm{W}$, the primal and dual residual norm are respectively denoted as $r$ and $s$ with mathematical expressions in (\ref{neADMM update of residual}), shown at the bottom of previous page.
It is obvious to see that the norms of the primal and dual residuals show an overall decreasing trend and gradually approach 0, which is consistent with the convergence behavior of the employed MM-FP-neADMM algorithmic framework \cite{neADMM 1}.
The three strategies based on the proposed algorithm in Sec. \ref{sec: beamforming} can converge within a hundred steps for a very high-dimensional non-convex problem.
More particularly, the ``Commun-only, Alg\ref{Algorithm Pseudocode}'' scheme without sensing constraint has the fastest convergence and achieves the highest sum-rate, while the ``W/o SE, Alg\ref{Algorithm Pseudocode}'' scheme yields the slowest convergence and the lowest sum-rate.
This phenomenon results from the varying degrees of sensing constraints imposed in these strategies.

\begin{figure}[!t]	
\centering
	\subfigure[Angle estimation performance]{
		\begin{minipage}[t]{1\linewidth}
			\centering\includegraphics[width=3.4 in]{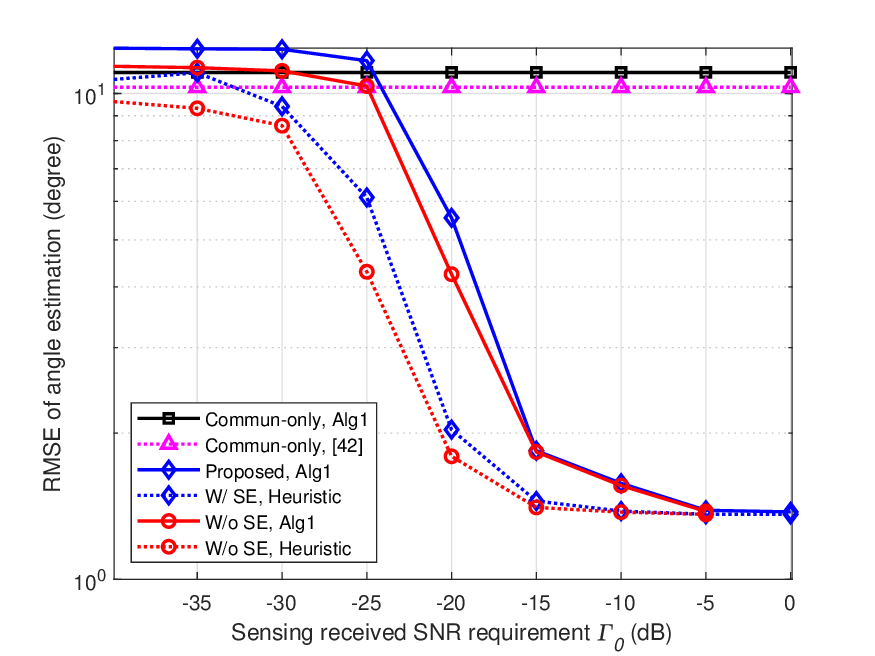}
		\end{minipage} \label{RMSEa}
	}\vspace{-2ex}

	\subfigure[Range estimation performance]{
		\begin{minipage}[t]{1\linewidth}
			\centering
			\includegraphics[width=3.4 in]{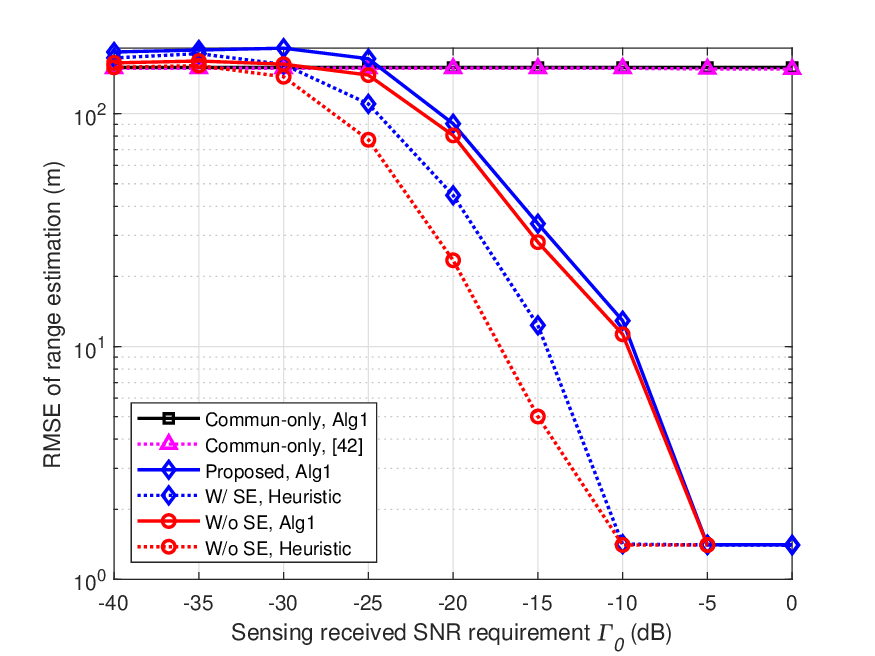}
		\end{minipage}\label{RMSEd}
	}\vspace{-2ex}

	\subfigure[Velocity estimation performance]{
		\begin{minipage}[t]{1\linewidth}
			\centering
			\includegraphics[width=3.4 in]{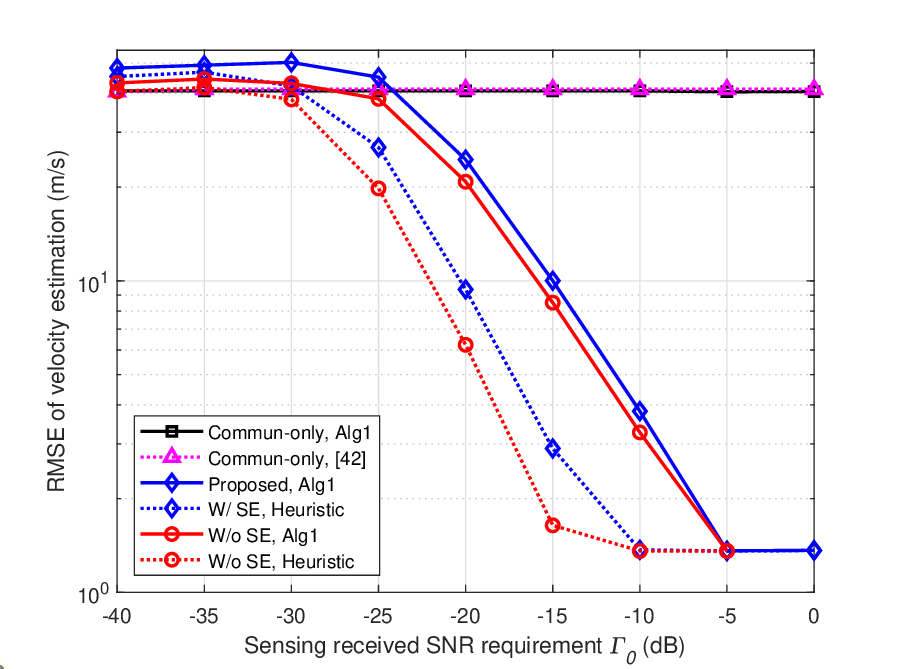}
		\end{minipage} \label{RMSEv}
	}\vspace{-1ex}
	\centering 
	\caption{Estimation performance versus sensing received SNR requirement $\Gamma_0$ ($P_0=10\mathrm{W}$).}
	\label{fig:RMSE} \vspace{-4ex}
\end{figure}


Next, we illustrate the average sum-rate as a function of sensing receive SNR requirement $\Gamma_\mathrm{0}$ in Fig. \ref{rate SNR fig}, where the total transmit power is $P_0 = 10\mathrm{W}$.
Due to the trade-off between sensing and communications, it can be noticed that the sum-rate of ISAC systems decreases as the sensing requirement threshold increases.
Consistent with the previously observed results in Fig. \ref{iteration fig}, we see that the beamforming design ``Proposed, Alg\ref{Algorithm Pseudocode}'' can achieve a higher rate than ``W/o SE, Alg\ref{Algorithm Pseudocode}''.
These results validate the advancement of the proposed sparsity exploitation strategy in reducing sensing expenses and thereby allocating more resources for communication usage. 
In constrast, the scheme without sparsity exploitation exhibits worse sum-rate performance, even cannot find a feasible solution with a relatively high sensing requirement as $\Gamma_\mathrm{0}=0\mathrm{dB}$, thereby no relevant simulation result being presented.
This advantage appears more obvious as the sensing requirement increases, with about a 50\% improvement at $\Gamma_0=-5\mathrm{dB}$.
In other words, the proposed strategy can be more advantageous in the high-sensing demanding scenarios.
For the same reason, the rate of ``W/ SE, Heuristic'' is larger than ``W/o SE, Heuristic''.
Meanwhile, the sum-rate of the proposed algorithm based designs is significantly higher than the heuristic algorithm based ones.
Different from the heuristic method, which performs dual-functional beamforming separately, the proposed algorithm treats it as a joint design, with which better interference management between sensing and communications can be achieved.
We also see that the sum-rate of the ``Commun-only, Alg\ref{Algorithm Pseudocode}'' strategy has comparable performance with the state-of-the-art WMMSE algorithm, which validates the effectiveness of our proposed algorithm in maximizing communication performance.

\begin{figure}[!t]
\centering
\includegraphics[width = 3.4 in]{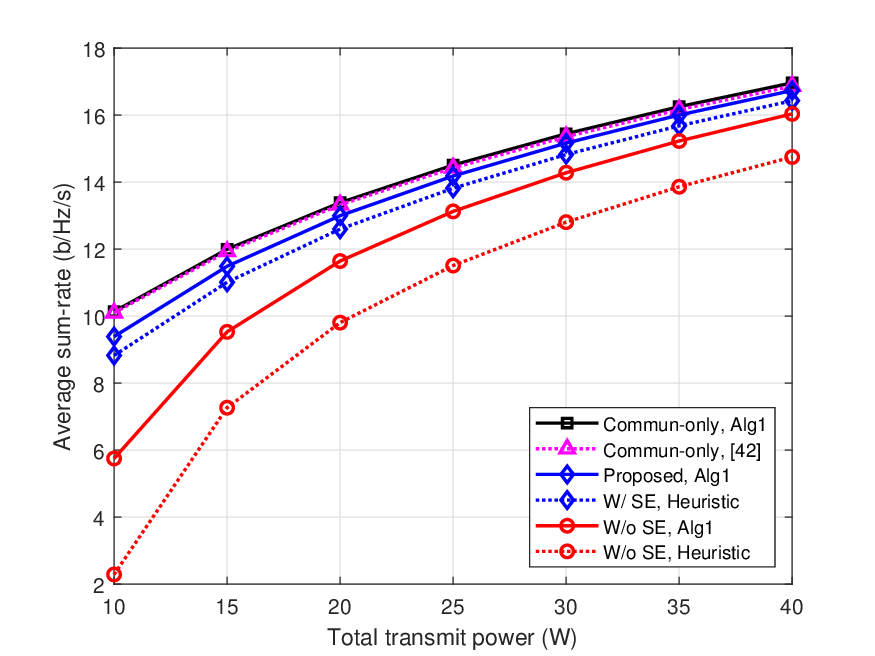}
\caption{Average sum-rate versus the transmit power $P_0$ (sensing received SNR requirement $\Gamma_0=-5 \mathrm{dB}$).} \vspace{-3 ex}
\label{fig: rate_P}
\end{figure}

In order to better evaluate the advantages and effectiveness of the proposed sparsity exploitation strategy, in Fig. \ref{fig:RMSE} we plot the corresponding angle-range-velocity estimation performance of the simulation in Fig. \ref{rate SNR fig} in terms of root-mean-squared-error (RMSE), where the targets are randomly distributed in the interested sensing region.
It can be noted that the RMSE of ISAC systems decreases as the sensing requirement threshold increases, which confirms the effectiveness of the suggested sensing SNR metric.
Meanwhile, the sensing performance of the proposed ``Proposed, Alg1'' strategy is nearly the same as that of the strategy ``W/o SE, Alg1''.
Accompanied by the communication performance illustrated in Fig. \ref{rate SNR fig}, it can be concluded that the proposed sparsity exploitation strategy with the proposed receive processing and transmit beamforming design can significantly improve the ISAC system efficiency.
Moreover, it can be observed that the heuristic algorithm shows a lower RMSE due to its tendency to over-fulfill the sensing demand with a loss of communication performance.
This phenomenon serves as additional evidence that our proposed algorithm has the capability to offer a more precise and manageable balance between sensing and communication performance.

Then, we illustrate the average sum-rate as a function of the total transmit power in Fig. \ref{fig: rate_P}. Clearly, a higher transmit power provides a larger sum-rate.
Moreover, the performance relationship is the same as shown in Fig. \ref{rate SNR fig} and similar conclusions can be drawn. 
One noteworthy phenomenon from this figure is that the performance gap between different schemes becomes smaller as the power budget increases.
This is because when the transmit power increases with a fixed sensing requirement, the sensing impact gradually becomes a fringe factor that affects the sum-rate performance.
Accordingly, Fig. \ref{fig: rate_P} reveals that the proposed sparsity exploitation strategy is more critical for tight power resources.


\begin{figure}[!t]
\centering
\includegraphics[width = 3.4 in]{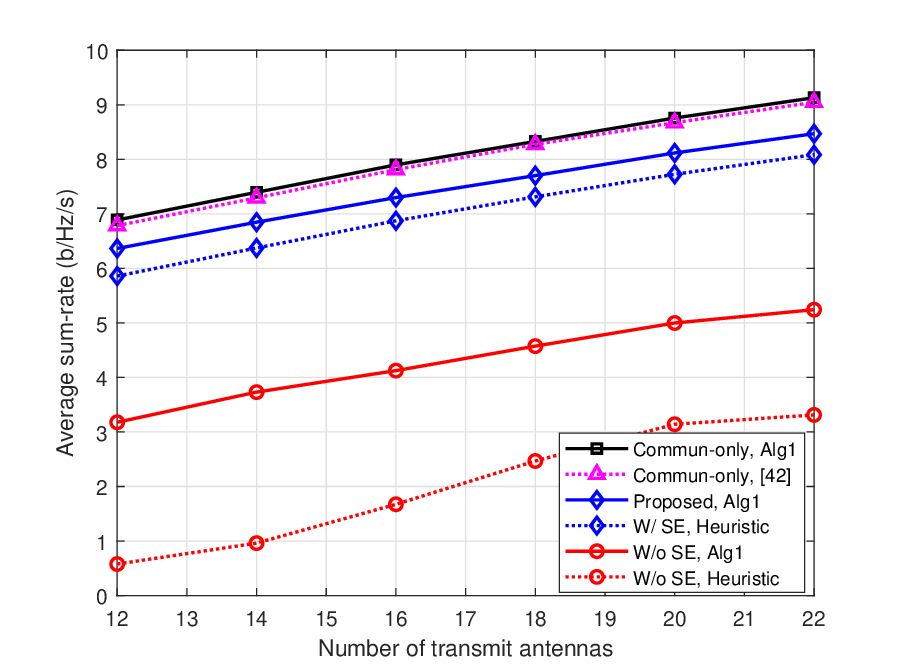}
\caption{Average sum-rate versus the number of transmit antennas $N_\mathrm{t}$ ($\Gamma_0=-5 \mathrm{dB}$).} \vspace{-3 ex}
 \label{fig: rate_Nt}
\end{figure}

\begin{figure}[!]
\centering
\includegraphics[width = 3.4 in]{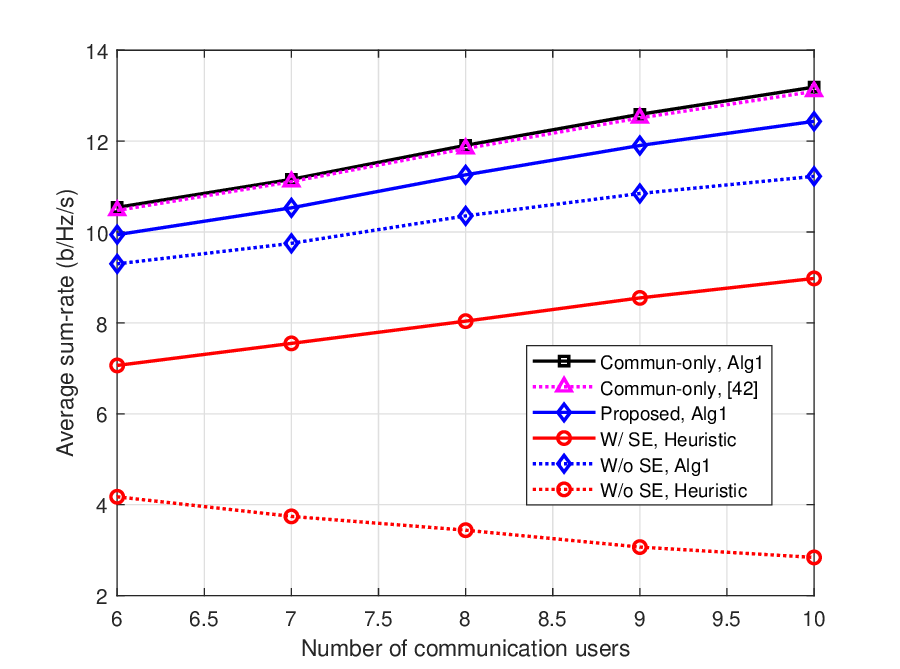}
\caption{Average sum-rate versus the number of communication users $K$ ($\Gamma_0=-5 \mathrm{dB}$).} \vspace{-3 ex}
 \label{fig: rate_K}
\end{figure}

In Fig. \ref{fig: rate_Nt}, we plot the communication sum-rate versus the number of transmit antennas.
It is evident that the system equipped with a greater number of antennas attains superior performance owing to its enhanced capability to suppress MUI and achieve higher beamforming gains.
A similar conclusion can be obtained from the above simulation results that the proposed scheme consistently outperforms its counterparts in terms of sum-rate performance.
Fig. \ref{fig: rate_K} illustrates the communication sum-rate as a function of the number of users.
Considering more users lead to larger DoFs for the systems to improve spectrum efficiency, we see that the sum-rate increases with the larger number of users except for scheme ``W/o, Heuristic''.
The reason for this exception lies in that the dual-functional beamforming in the heuristic algorithm is calculated in a separate way without interference management, with which the interference will be one dominant factor affecting the sum-rate. 
More importantly, under different antenna array sizes or user populations, the proposed sparsity exploitation strategy and the associated beamforming design algorithm can always provide the ISAC system with superior sum-rate performance close to the communication-only system. 


\section{Conclusions}  
In this paper, we investigated the sparsity exploitation via joint receive processing and transmit beamforming design for the MIMO-OFDM ISAC system.
One CS-based estimation method was first proposed for reducing the required frequency expenses for sensing.
Then, dedicated beamforming was designed to maximize the communication sum-rate under the constraint that guarantees reliable sensing receiving of interested region and utilized frequency subcarriers.
Efficient algorithms exploiting MM, FP, and neADMM methods were developed to solve the resulting high-dimensional non-convex optimization problems.
Simulation examples demonstrated the effectiveness of the proposed algorithm and the advantages of the proposed sparsity exploitation strategy.

\appendices
\section{Proof of Proposition 1}
To clarify whether the real-valued minimization problem (\ref{neADMM update of W mao}) is bounded below, it is equivalent to examining whether the complex-valued minimization (\ref{neADMM update of W}) is bounded below.
We observe that the objective function of optimization (\ref{neADMM update of W}) has two non-negative norm terms, which can therefore be dropped in the following analysis.
Thus, this examination is simplified to figure out whether the left term $f_i(\mathbf{w}_i)$ is bounded below.
Detailed analysis is provided in the following.

Since function $f_i(\mathbf{w}_i)$ is quadratic convex, the minimizer of it should have
\begin{equation}
\begin{aligned}
\frac{\partial f_i(\mathbf{w}_i)}{ \partial \mathbf{w}_i^*}
=\mathbf{T}_i\mathbf{w}_i-\widehat{\mathbf{h}}_i=\mathbf{0}.\label{linear equation of fi}
\end{aligned}
\end{equation}
Substituting the expressions (\ref{vec w definition}) and (\ref{T i definitions}) into the above equation (\ref{linear equation of fi}) , it can be converted into $K$ sub-equations as
\begin{equation}
\begin{aligned}
\mathbf{H}_i \mathbf{\Lambda}_i \mathbf{H}_i^H \mathbf{w}_{i,k}= \sqrt{1+r_{i,k}}t_{i,k}\mathbf{h}_{i,k}, ~~\forall k, \label{subequations 2}
\end{aligned}
\end{equation}
where the matrix $\mathbf{\Lambda}_i$ is defined as $\mathbf{\Lambda}_i\triangleq \mathrm{Diag}([|t_{i,1}|^2, \ldots, |t_{i,K}|^2]^T)$.
Based on the assumption (A2) and basic linear algebra knowledge, equation (\ref{subequations 2}) has the following equivalent equation
\begin{equation}
\begin{aligned}
\mathbf{\Lambda}_i \mathbf{H}_i^H \mathbf{w}_{i,k}= \sqrt{1+r_{i,k}}t_{i,k} \mathbf{e}_k, ~~\forall k, \label{subequations 3}
\end{aligned}
\end{equation}
where $\mathbf{e}_k\triangleq \big[[0,\ldots,0]_{1\times k-1},1,[0,\ldots,0]_{1\times K-k}\big]^T$.
With assumption (A2), $\mathbf{\Lambda}_i \mathbf{H}_i^H$ has full row rank, thereby the optimal solution of equation (\ref{subequations 3}) exists and one of them is $\sqrt{1+r_{i,k}}t_{i,k} (\mathbf{\Lambda}_i \mathbf{H}_i^H)^\dag \mathbf{e}_k$.
Therefore, we have proved that the minimizer of the function $f_i(\mathbf{w}_i)$ exists and function $f_i(\mathbf{w}_i)$ is bounded below, thereby relevant real-valued optimization (\ref{neADMM update of W mao}) is bounded below.

Since optimization (\ref{neADMM update of W mao}) is bounded below, the solution of equation (\ref{linear equation of W}) exists and one is
\be
\widehat{\mathbf{w}}_i^{\star}\triangleq (\mathbf{A}_i^{(t)}+{\mathbf{A}_i^{(t)}}^T)^{\dag}\mathbf{b}_i^{(t)},
\ee
which completes the proof.


\begin{thebibliography}{99}
\bibitem{ITU future prospective} ITU-R, DRAFT NEW RECOMMENDATION, ``Framework and overall objectives of the future development of IMT for 2030 and beyond,'' Jun. 2023.
\bibitem{Andrew Zhang overview} J. A. Zhang \textit{et al.}, ``Enabling joint communication and radar sensing in mobile networks - A survey,'' \textit{IEEE Commun. Surveys Tuts.}, vol. 24, no. 1, pp. 306-345, 1st Quart., 2022.
\bibitem{F Liu overview} F. Liu \textit{et al.}, ``Integrated sensing and communications: Toward dual-functional wireless networks for 6G and beyond,'' \textit{IEEE J. Sel. Areas Commun.}, vol. 40, no. 6, pp. 1728-1767, Jun. 2022.
\bibitem{Chirp based 1} M. Roberton and E. R. Brown, ``Integrated radar and communications based on chirped spread-spectrum techniques,'' in \textit{IEEE MTT-S Int. Microw. Symp. Dig.}, Philadelphia, USA, Jun. 2003, pp. 611-614.
\bibitem{FH radar based 2} K. Wu, J. A. Zhang, X. Huang, and Y. J. Guo, ``Frequency-hopping MIMO radar-based communications: An overview,'' \textit{IEEE Aerosp. Electron. Syst. Mag.}, vol. 37, no. 4, pp. 42-54, Apr. 2022.
\bibitem{FAR radar based} T. Huang, N. Shlezinger, X. Xu, Y. Liu, and Y. C. Eldar, ``MAJoRCom: A dual-function radar communication system using index modulation,'' \textit{IEEE Trans. Signal Process.}, vol. 68, pp. 3423-3438, May 2020.
\bibitem{C Sturm} C. Sturm and W. Wiesbeck, ``Waveform design and signal processing aspects for fusion of wireless communications and radar sensing,'' \textit{IEEE Proc.}, vol. 99, no. 7, pp. 1236-1259, Jul. 2011.
\bibitem{80211 1} P. Kumari, J. Choi, N. G. Prelcic, and R. W. Heath, ``IEEE 802.11ad-based radar: An approach to joint vehicular communication-radar system,'' \textit{IEEE Trans. Veh. Technol.}, vol. 67, no. 4, pp. 3012-3027, Apr. 2018.
\bibitem{OTFS ISAC} K. Wu, J. A. Zhang, X. Huang, and Y. J. Guo, ``Integrating low-complexity and flexible sensing into communication systems,'' \textit{IEEE J. Sel. Topics Signal Process.}, vol. 40, no. 6, pp. 1873-1889, Jun. 2022.
\bibitem{RIS ISAC} R. Liu, M. Li, Y. Liu, Q. Wu, and Q. Liu, ``Joint transmit waveform and passive beamforming design for RIS-aided DFRC systems,'' \textit{IEEE J. Sel. Topics Signal Process.}, vol. 16, no. 5, pp. 995-1010, Aug. 2022.
\bibitem{beampattern error and rate} Z. Cheng, Z. He, and B. Liao, ``Hybrid beamforming design for OFDM dual-function radar-communication system,'' \textit{IEEE J. Sel. Topics Signal Process.}, vol. 15, no. 6, pp. 1455-1467, Nov. 2021.
\bibitem{beampattern error and MUI} X. Hu, C. Masouros, F. Liu, and Ronald Nissel, ``MIMO-OFDM dual-functional radar-communication systems: Low-PAPR waveform design,'' Sep. 2021. [Online]. Available: https://arxiv.org/abs/2109.13148
\bibitem{SINR and SINR 1} T. Wei, L. Wu, K. V. Mishra, and M. R. B. Shankar, ``Multiple IRS-assisted wideband dual-function radar-communication,'' in \textit{Proc. 2022 2nd IEEE Int. Symp. Joint Commun. \& Sens. (JC\&S)}, Seefeld, Austria, Mar. 2022, pp. 1-5.
\bibitem{SINR and SINR 2} Z. Xiao, R. Liu, M. Li, Y. Liu, and Q. Liu, ``Joint beamforming design in DFRC systems for wideband sensing and OFDM communications,'' in \textit{Proc. IEEE Global Commun. Conf. (GLOBECOM)}, Rio, Brazil, Dec. 2022, pp. 1631-1636.
\bibitem{SINR and QoS} J. Johnston, L. Venturino, E. Grossi, M. Lops, and X. Wang, ``MIMO OFDM dual-function radar-communication under error rate and beampattern constraints,'' \textit{IEEE J. Sel. Areas Commun.}, vol. 40, no. 6, pp. 1951-1964, Jun. 2022.
\bibitem{CS-ISAC Rahman1} M. L. Rahman, P.-F. Cui, J. A. Zhang, X. Huang, Y. J. Guo, and Z. Lu, ``Joint communication and radar sensing in 5G mobile network by compressive sensing,'' in \textit{Proc. 19th Int. Symp. Commun. Inf. Technol. (ISCIT)}, Ho Chi Minh City, Vietnam, Sep. 2019, pp. 599-604.
\bibitem{CS-ISAC Rahman2} M. L. Rahman, J. A. Zhang, X. Huang, Y. J. Guo, and R. W. Heath, ``Framework for a perceptive mobile network using joint communication and radar sensing,'' \textit{IEEE Trans. Aerosp. Electron. Syst.}, vol. 56, no. 3, pp. 1926-1941, Jun. 2020.
\bibitem{MIMO OFDM ISAC successive estimation 1} M. A. Islam, G. C. Alexandropoulos, and B. Smida, ``Integrated sensing and communication with millimeter wave full duplex hybrid beamforming,'' in \textit{Proc. IEEE Int. Conf. Commun. (ICC)}, Seoul, Korea, May 2022, pp. 4673-4678.
\bibitem{MIMO OFDM ISAC successive estimation 2} Z. Xu and A. Petropulu, ``A bandwidth efficient dual-function radar communication system based on a MIMO radar using OFDM waveforms,'' \textit{IEEE Trans. Signal Process.}, vol. 71, pp. 401-416, Feb. 2023.
\bibitem{our estimation work} Z. Xiao, R. Liu, M. Li, and Q. Liu, ``A novel joint angle-range-velocity estimation method for MIMO-OFDM ISAC systems,'' Aug. 2023. [Online]. Available: https://arxiv.org/abs/2308.03387
\bibitem{CS theory 1} D. V. Donoho, ``Compressed sensing,'' \textit{IEEE Trans. Inf. Theory}, vol. 52, no. 4, pp. 1289-1306, Apr. 2006.
\bibitem{CS aided sensing (reduce cost)1} Y. Yu, A. P. Petropulu, and H. V. Poor, ``MIMO radar using compressive sampling,'' \textit{IEEE J. Sel. Topics Signal Process.}, vol. 4, no. 1, pp. 146-163, Feb. 2010.
\bibitem{CS aided sensing (reduce cost)3} D. Cohen, D. Cohen, Y. C. Eldar, and A. M. Haimovich, ``SUMMeR: Sub-Nyquist MIMO radar,'' \textit{IEEE Trans. Signal Process.}, vol. 66, no. 16, pp. 4315-4330, Aug. 2018.
\bibitem{CS aided sensing (enhance resolution)1} M. A. Herman and T. Strohmer, ``High-resolution radar via compressed sensing,'' \textit{IEEE Trans. Signal Process.}, vol. 57, no. 6, pp. 2275-2284, Jun. 2009.
\bibitem{CS aided sensing (enhance detection)} A. Ajorloo, A. Amini, and M. H. Bastani, ``A compressive sensing-based colocated MIMO radar power allocation and waveform design,'' \textit{IEEE Sensors J.}, vol. 18, no. 22, pp. 9420-9429, Nov. 2018.
\bibitem{CS aided communications1} C. R. Berger, Z. Wang, J. Huang, and S. Zhou, ``Application of compressive sensing to sparse channel estimation,'' \textit{IEEE Commun. Mag.}, vol. 48, no. 11, pp. 164-174, Nov. 2010.
\bibitem{CS aided communications2} X. Ma, F. Yang, S. Liu, J. Song, and Z. Han, ``Design and optimization on training sequence for mmWave communications: A new approach for sparse channel estimation in massive MIMO,'' \textit{IEEE J. Sel. Areas Commun.}, vol. 35, no. 7, pp. 1486-1497, Jul. 2017.
\bibitem{CS aided ISAC} Z. Gao, Z. Wan, D. Zheng, S. Tan, C. Masouros, D. W. K. Ng, and S. Chen, ``Integrated sensing and communication with mmWave massive MIMO: A compressed sampling perspective,'' \textit{IEEE Trans. Wireless Commun.}, vol. 22, no. 3, pp. 1745-1762, Mar. 2023.
\bibitem{channel estimation 2} Y. Liu, Z. Tan, H. Hu, L. J. Cimini, and G. Y. Li, ``Channel estimation for OFDM,'' \textit{IEEE Commun. Surveys Tuts.}, vol. 16, no. 4, pp. 1891-1908, 4th Quart., 2014.
\bibitem{book: Emil} E. Bj\"{o}rnson, J. Hoydis, and L. Sanguinetti, ``Massive MIMO networks: Spectral, energy, and hardware efficiency,'' \textit{Found. Trends Signal Process.}, vol. 11, no. 3-4, pp. 154-655, 2017.
\bibitem{CS optimization} J. A. Tropp and S. J. Wright, ``Computational methods for sparse solution of linear inverse problems,'' \textit{IEEE Proc.}, vol. 98, no. 6, pp. 948-958, Jun. 2010.
\bibitem{CS RIP 1} E. J. Cand\`{e}s, J. Romberg, and T. Tao, ``Robust uncertainty principles: Exact signal reconstruction from highly incomplete frequency information,'' \textit{IEEE Trans. Inf. Theory}, vol. 52, no. 2, pp. 489-509, Feb. 2006.
\bibitem{CS RIP 2} S. Foucart and M.-J. Lai, ``Sparsest solutions of underdetermined linear systems via $\ell_q$-minimization for $0<q\leq1$,'' \textit{Appl. Comput. Harmonic Anal.}, vol. 26, no. 3, pp. 395-407, 2009.
\bibitem{MM 2} Y. Sun, P. Babu, and D. P. Palomar, ``Majorization-minimization algorithms in signal processing, communications, and machine learning,'' \textit{IEEE Trans. Signal Process.}, vol. 65, no. 3, pp. 794-816, Feb. 2017.
\bibitem{FP 1} H. Li, M. Li, and Q. Liu, ``Hybrid beamforming with dynamic subarrays and low-resolution PSs for mmWave MU-MISO systems,'' \textit{IEEE Trans. Commun.}, vol. 68, no. 1, pp. 602-614, Jan. 2020.
\bibitem{FP 2} K. Shen and W. Yu, ``Fractional programming for communication systems-Part I: Power control and beamforming,'' \textit{IEEE Trans. Signal Process.}, vol. 66, no. 10, pp. 2616-2630, May 2018.
\bibitem{neADMM 1} J. Wang and L. Zhao, ``Nonconvex generalization of alternating direction method of multipliers for nonlinear equality constrained problems,'' \textit{Results Control Optim.}, vol. 2, Mar. 2021, Art. no. 100009.
\bibitem{book: Boyd} S. Boyd and L. Vandenberghe, \textit{Convex Optimization}. Cambridge, U.K.: Cambridge Univ. Press, 2004.
\bibitem{cubic formula} Wikipedia contributors. ``Cubic equation.'' \textit{Wikipedia, The Free Encyclopedia}, Apr. 2023. [Online]. Available: https://en.wikipedia.org/w/index.php?title=Cubic\_equation\&oldid=1149\\948974
\bibitem{book: active set} J. Nocedal and S. Wright, \textit{Numerical Optimization}. Berlin, Germany: SpringerVerlag, 2006.
\bibitem{3GPP 211} 3GPP, ``3GPP TS 38.211 V17.4.0 (2022-12) 3rd Generation Partnership Project; Technical Specification Group Radio Access Network; NR; Physical channels and modulation (Release 17),'' 2022.
\bibitem{WMMSE} Q. Shi, M. Razaviyayn, Z.-Q. Luo, and C. He, ``An iteratively weighted MMSE approach to distributed sum-utility maximization for a MIMO interfering broadcast channel,'' \textit{IEEE Trans. Signal Process.}, vol. 59, no. 9, pp. 4331-4340, Sep. 2011.


\end{thebibliography}
\end{document}